# Computing the Expected Value and Variance of Geometric Measures

*Frank Staals* *Constantinos Tsirogiannis*
*MADALGO*[*]
*Aarhus University, Denmark*
`[f.staals|constant]@cs.au.dk`

#### Abstract

Let $P$ be a set of points in $\mathbb{R}^d$, and let $M$ be a function that maps any subset of $P$ to a positive real number. We examine the problem of computing the exact mean and variance of $M$ when a subset of points in $P$ is selected according to a well-defined random distribution. We consider two distributions; in the first distribution (which we call the *Bernoulli* distribution), each point $p \in P$ is included in the random subset independently, with probability $\pi(p)$. In the second distribution (the *fixed-size* distribution), a subset of exactly $s$ points is selected uniformly at random among all possible subsets of $s$ points in $P$.

This problem is a crucial part of modern ecological analyses; each point in $P$ represents a species in $d$-dimensional trait space, and the goal is to compute the statistics of a geometric measure on this trait space, when subsets of species are selected under random processes.

We present efficient exact algorithms for computing the mean and variance of several geometric measures when point sets are selected under one of the described random distributions. More specifically, we provide algorithms for the following measures: the bounding box volume, the convex hull volume, the mean pairwise distance (MPD), the squared Euclidean distance from the centroid, and the diameter of the minimum enclosing disk. We also describe an efficient $(1-\varepsilon)$-approximation algorithm for computing the mean and variance of the mean pairwise distance.

We implemented three of our algorithms: an algorithm that computes the exact mean volume of the 2D bounding box in the Bernoulli distribution, an algorithm that computes the exact mean and variance of the MPD for $d$-dimensional point sets in the fixed-size distribution, and an $(1-\varepsilon)$-approximation algorithm for the same measure. We conducted experiments where we compared the performance of our implementations with a standard heuristic approach used in ecological applications. We show that our implementations can provide major speedups compared to the standard approach, and they produce results of higher precision, especially for the calculation of the variance. We also compared the implementation of our exact MPD algorithm with the corresponding $(1-\varepsilon)$-approximation method; we show that the approximation method performs faster in certain cases, while also providing high-precision approximations. We thus demonstrate that, as an alternative to the exact algorithm, this method can also be used as a reliable tool for ecological analysis.

## 1 Introduction

**Motivation.** Scientists in Ecology are devoted to the study of ecosystems and the processes that make these systems viable. Among other properties, ecologists are interested in measuring the *biodiversity* of an ecosystem: the diversity of the species that live inside the ecosystem. There are different ways to express this diversity, and therefore different measures for evaluating it. These measures are functions that map the set of species living in the ecosystem to a positive real number. One of the most important categories of diversity measures are *functional diversity* measures; these measures evaluate the diversity of functional traits that are observed within a set of species[18, 15, 20]. More formally, let $S$ be the set of species that appear in an ecosystem that we want to examine. For each species in $S$, ecologists measure the values of $d$ different traits e.g. body mass, the latitude where this species is commonly observed etc.. In this way, each species in $S$ can be represented as a point in $d$-dimensional space. For simplicity, from hereon we use $S$ to denote the $d$-dimensional point set representing the species of the ecosystem. Given the point set $S$, a functional diversity measure $M$ is a real-valued function which measures a geometric property of $S$. For example, one of the most popular functional measures used by ecologists is the volume of the convex hull of $S$ [5]. Another frequently used measure is the so-called *Mean Pairwise Distance* (MPD), which is equal to the average Euclidean distance among all distinct pairs of points in $S$ [21].

Whichever measure $M$ is used, in most ecological applications it is not enough just to compute the value $M(S)$ on the examined point set $S$. It is also important to determine if $M(S)$ is significantly larger or smaller than the value of this measure for a randomly selected set of species. The result of this comparison could indicate if there is a special reason why the species in $S$ appear together in the same ecosystem, or their co-existence resembles the result of a random process. More specifically, let $P \supseteq S$ be a point set in $\mathbb{R}^d$ representing a universal pool of species that we want to consider. To measure the significance of the value of $M$ for a specific set $S$, ecologists want to estimate the distribution of $M(S')$, where $S'$ is a subset of $P$ selected according to a random distribution. To do this, they usually calculate the expected value and the variance of $M(S')$. Based on these two values, they can then decide if the observed set of species $S$ is special, with respect to $M$, compared to a set resulting from a random process.

To calculate the expected value and variance of $M(S')$, we first need to define the distribution based on which we select a random point set $S'$. We consider two of the most popular distributions that appear in ecological applications: the *Bernoulli* distribution and the *fixed-size* distribution. In the Bernoulli distribution, each point $p \in P$ is associated with a probability value $\pi(p)$; this value represents the abundance of the corresponding species in the real world. To produce a random set $S'$ according to the Bernoulli distribution, each point $p$ is selected for inclusion to $S'$ by performing an independent Bernoulli trial with probability of success $\pi(p)$. In the fixed-size distribution a subset of exactly $s$ points is selected from $P$, and all possible subsets of $s$ species can be selected with equal probability. The size $s$ of the subset equals the number of species of the observed community $S$ that is examined. Note that, unlike in the Bernoulli distribution, the selection of each point in the fixed-size distribution is not statistically independent.

Computing the statistics of a geometric measure $M$ for the above random distributions can be a hard computational task. So far, ecologists have been calculating these statistics using a crude heuristic approach; first, a large number of point samples (typically, at least a thousand) is produced based on one of the above distributions. Then, the value of the examined measure $M$ is calculated for each of these samples, and finally an estimation of the mean and the variance of $M$ is derived based on these calculations.

This heuristic approach has two main disadvantages. First, it does not provide any approximation guarantee between the calculated mean and variance and the actual statistics of $M$ for the chosen random distribution. Second, it is often very slow in practice since it requires to compute the value of $M$ for a large number of samples. Hence, this crude method is a major obstacle for conducting reliable ecological analyses, let alone to process a large amount of species data. Therefore, there is the need to design efficient algorithms which can calculate exactly the expected value and variance of standard geometric functions over random point set distributions. More than that, it is important to derive robust implementations of these algorithms, that perform very fast when applied on real ecological datasets. Such implementations would improve the quality of the analysis that is presented in ecological case studies, and thus lead to more accurate conclusions on the properties of ecosystems.

**Our Results.** We present algorithms that compute the exact statistical moments of several geometric functions on random point sets selected either under the Bernoulli or the fixed-size distribution. In particular, given a set of $n$ input points we describe:

- An $O(n \log n)$ time algorithm that computes the expected bounding-box volume for a set of points in $\mathbb{R}^2$, selected under the Bernoulli distribution (Section 3). The algorithm can be extended to compute the mean bounding-box volume for points in $\mathbb{R}^d$ in $O(n^{d-1} \log n)$ time.
- An $O(n^d \log n)$ time algorithm that computes the mean of the convex hull volume for a set of points in $\mathbb{R}^d$ selected either under the Bernoulli or the fixed-size distribution (Section 4).
- An $O(n)$ time algorithm that computes the mean and variance for the squared Euclidean distance of the points in a subset $S \subset \mathbb{R}^d$ from the centroid of $S$, when $S$ is selected under the fixed-size distribution (Section 5).
- An $O(n \log n + n/\varepsilon^d)$ time $(1-\varepsilon)$-approximation algorithm that computes the mean and variance of the mean pairwise distance for a set of points in $\mathbb{R}^d$ selected under the fixed-size distribution (Section 6). These statistics can be computed exactly in $O(n^2)$ time.
- An $O(n^3 \log n)$ time algorithm that computes the mean diameter of the smallest enclosing disk for a set of points in $\mathbb{R}^3$, selected under either distribution (Section 7).

We implemented some of the above algorithms, and evaluated their performance on both artificial and real species data (Section 8). More specifically, we implemented the algorithm that computes the mean volume of the 2D bounding box, and the exact and approximate algorithms that compute the mean and variance of the MPD. We compare our algorithms with the heuristic sampling method currently used by ecologists. These experiments show that our implementations can be much faster than the sampling approach, while providing guarantees on the quality of the resulting output. Our implementations were developed in C++ and are publicly available through github [19].

**Related Work.** There have been several papers in Algorithms that study geometric structures on stochastic point sets [1, 6, 17], yet most of these do not examine the exact computation of the statistical moments of geometric measures. Jørgensen et al. [10] study the problem of computing the distribution of geometric measures on a finite set of points, where each point can be in one out of $k$ given positions with certain probability. Löffler and Phillips [13] present algorithms that approximate the distributions of answers for problems like the minimum enclosing ball radius, when the location of the input points is provided in the form of a distribution. Li et al. [12] study the existence and extraction of core sets to approximate the expected diameter of a sample of points selected under a random distribution. One of the distributions they consider is the Bernoulli distribution (which they refer to as the "existential model"). Huang and Li [9] present approximation algorithms for several related problems, some of which are known to be #P-hard.

## 2 Definitions and Notation

Let $P$ be a set of $n$ points in $\mathbb{R}^d$, where $d$ is a constant. For ease of description, we assume that $P$ is in general position; no pair of these points share the same coordinate, and for any natural number $k \leq d$ there does not exist any subset of $k+1$ points in $P$ that lie on the same $(k-1)$-dimensional hyperplane. Let $M$ be a function that maps any set $S$ of points in $P$ to a positive real number; for example, $M(S)$ is the volume of the convex hull of $S$. We call such a function a *measure* of $S$. Consider that we select a subset $S$ from $P$ based on a given random distribution. We study the problem of computing the exact expected value $\mathbb{E}[M(S)]$ and variance $\mathbb{V}[M(S)]$ for several measures $M$ of $S$. The expectation of $M(S)$ is defined as $\mathbb{E}[M(S)] = \sum_{Q \subseteq P} M(Q) \cdot \mathbb{P}[S = Q]$, where $S$ is a random variable, and $\mathbb{P}[S = Q]$ is the probability that the subset of points $Q$ is selected according to the described distribution. The variance of $M(S)$ is equal to $\mathbb{E}[M^2(S)] - \mathbb{E}[M(S)]^2$. In the Bernoulli distribution, each point in $P$ is associated with a probability value. For a point $p$ in $P$, we denote this value by $\pi(p)$. Let $Q$ be a subset of $P$. We use $\Pi(Q) = \prod_{p \in Q} \pi(p)$ to denote the probability of selecting all points in the set $Q$, and $\overline{\Pi}(Q) = \prod_{p \in Q}(1 - \pi(p))$ to denote the probability of not selecting *any* point in $Q$.

## 3 The Volume of the Bounding Box

Let $S$ be a subset of points in $P$, and let $\mathcal{BB}(S)$ denote the volume of the bounding box of $S$. In this section we describe efficient algorithms for computing the expected value of $\mathcal{BB}(S)$ when $S$ is selected under the Bernoulli distribution. We begin with the case where $P \subset \mathbb{R}^2$; for this case, we present an algorithm that runs in $O(n \log n)$ time. Let $p$ be a point in $\mathbb{R}^2$. We use $p_x$ and $p_y$ to denote the $x$ and $y$ coordinates of $p$, respectively. We have that

$$\mathbb{E}[\mathrm{vol}(\mathcal{BB}(S))] = \mathbb{E}\left[\left(\max_{p \in S} p_x - \min_{q \in S} q_x\right)\left(\max_{r \in S} r_y - \min_{t \in S} t_y\right)\right]$$
$$= \mathbb{E}\left[\max_{p \in S} p_x \cdot \max_{r \in S} r_y\right] - \mathbb{E}\left[\max_{p \in S} p_x \cdot \min_{t \in S} t_y\right] - \mathbb{E}\left[\min_{q \in S} q_x \cdot \max_{r \in S} r_y\right] + \mathbb{E}\left[\min_{q \in S} q_x \cdot \min_{t \in S} t_y\right]. \quad (1)$$

We focus on computing the term $\mathbb{E}[\max_{p \in S} p_x \cdot \max_{q \in S} q_y]$. The other three terms can be computed in a similar manner. We have

$$\mathbb{E}\left[\max_{p \in S} p_x \cdot \max_{q \in S} q_y\right] = \sum_{p,q \in P} p_x q_y \cdot \mathbb{P}\left[p_x = \max_{r \in S} r_x \text{ and } q_y = \max_{t \in S} t_y\right]. \quad (2)$$

Let $P_x^+(p)$ be the points that have a larger $x$-coordinate than $p$. Similarly, let $P_y^+(p)$ be the points with a larger $y$-coordinate than $p$. We also use $P^+(p, q)$ to denote the set of points that have both a larger $x$-coordinate than $p$ and a larger $y$-coordinate than $q$. The probability value in Eq. (2) is equal to the probability that $p$ and $q$ are selected in $S$, and there is no point in $S$ that has either a $x$-coordinate larger than $p_x$ or a $y$-coordinate larger than $q_y$. The set of points that violate the latter condition are the points in $P$ in $P_x^+(p) \cup P_y^+(q)$.

**Lemma 1.** *Let $S$ be a subset of $P$ selected according to the Bernoulli distribution. We have*

$$\mathbb{E}[\max_{p \in S} p_x \cdot \max_{q \in S} q_y] = \sum_{p \in P} p_x \cdot \pi(p) \cdot \overline{\Pi}(P_x^+(p)) \sum_{q \in P_x^-(p) \cup P_y^+(p)} q_y \cdot \pi(q) \cdot \overline{\Pi}(P_y^+(q)) \cdot \frac{1}{\overline{\Pi}(P^+(p,q))} +$$
$$\sum_{g \in P} g_x g_y \cdot \pi(g) \cdot \overline{\Pi}(P_x^+(g) \cup P_y^+(g)). \quad (3)$$

*Proof.* We consider two cases; in the first case, the point that has the maximum $x$-coordinate in $S$ is different than the point in $S$ with the maximum $y$-coordinate. In the second case, it is the same point in $S$ that has both the maximum $x$ and $y$ coordinates in this set. Based on this distinction, we can rewrite Eq. (2) as:

$$\mathbb{E}\left[\max_{p\in S} p_x \cdot \max_{q\in S} q_y\right] = \sum_{\substack{p,q\in P\\ p\neq q}} p_x q_y \cdot \mathbb{P}\left[p_x = \max_{r\in S} r_x \text{ and } q_y = \max_{t\in S} t_y\right] + \sum_{g\in P} g_x g_y \cdot \mathbb{P}\left[g_x = \max_{u\in S} u_x \text{ and } g_y = \max_{v\in S} v_y\right]. \tag{4}$$

We continue by expanding further the probability values that appear in the last equation. Let $p$ and $q$ be two distinct points in $P$. The probability that $p$ has the maximum $x$-coordinate and $q$ has the maximum $y$-coordinate in $S$ is zero if either $q \in P_x^+(p)$ or $p \in P_y^+(q)$. Otherwise, this probability is equal to the probability of selecting $p$ and $q$ times the probability of not selecting any point in $P_x^+(p) \cup P_y^+(q)$. In the latter situation, we get:

$$\mathbb{P}\left[p_x = \max_{r\in S} r_x \text{ and } q_y = \max_{t\in S} t_y\right] = \pi(p)\cdot\pi(q)\cdot\overline{\pi}(P_x^+(p)\cup P_y^+(q)) = \pi(p)\cdot\pi(q)\cdot\overline{\pi}(P_x^+(p))\cdot\overline{\pi}(P_y^+(q))\cdot\frac{1}{\overline{\pi}(P^+(p,q))}.$$

For a single point $p \in P$, the probability that $p$ has the maximum coordinate in $S$ both for $x$ and $y$ is:

$$\mathbb{P}\left[p_x = \max_{r\in S} r_x \text{ and } p_y = \max_{t\in S} t_y\right] = \pi(p)\cdot\overline{\pi}(P_x^+(p)\cup P_y^+(p)).$$

The lemma follows by combining the two last equations with Eq. (4). □

Our next step is to design an efficient method for evaluating the quantities that appear in the statement of Lemma 1. Given that, it is then straightforward to derive an efficient algorithm for computing the expected volume of the bounding box under the Bernoulli distribution. We continue by breaking the formula at the right side of Eq. (3) into simpler quantities. In particular, we can rewrite this formula to $\mathbb{E}[\max_{p\in S} p_x \cdot \max_{q\in S} q_y] = \sum_{p\in P}(A(p)\cdot B(p) + C(p))$, where

$$A(p) = p_x\cdot\pi(p)\cdot\overline{\pi}(P_x^+(p)), \qquad B(p) = \sum_{q\in P_x^-(p)\cup P_y^+(p)} q_y\cdot\pi(q)\cdot\overline{\pi}(P_y^+(q))\cdot\frac{1}{\overline{\pi}(P^+(p,q))},$$ and

$C(p) = p_x p_y\cdot\pi(p)\cdot\overline{\pi}(P_x^+(p)\cup P_y^+(p))$. We can compute $A(p)$ for every $p \in P$ in $O(n\log n)$ time in total in the following manner; we first sort the points in $P$ in decreasing $x$-coordinate, and then we calculate value $\overline{\pi}(P_x^+(q))$ for each point $q$ based on the corresponding quantity of its predecessor in this order. However, quantities $B(p)$ and $C(p)$ are more complicated, and we have to follow a more involved approach. Yet, we can prove that these values can be computed for all points in $P$ in $O(n\log n)$ time in total.

**Lemma 2.** *We can compute $B(p)$ and $C(p)$ for every $p \in P$ in $O(n \log n)$ time in total.*

*Proof.* We begin by describing a method for computing values $B(p)$ for every $p \in P$. We write $B(p)$ as

$$B(p) = \sum_{q \in P_x^-(p) \cup P_y^+(p)} D(q) \cdot \frac{1}{\overline{\overline{\pi}}(P^+(p,q))}\,, \tag{5}$$

where $D(q) = q_y \cdot \pi(q) \cdot \overline{\overline{\pi}}(P_y^+(q))$. We can then precompute all values $D(q)$, $q \in P$ in $O(n \log n)$ time in a similar way as we did with values $A(p)$. Given these values, we proceed with the computation of $B(p)$ for every $p \in P$ based on Eq. (5). To do this, we use a data structure $\mathcal{T}_{\text{prod}}$ that we call the *product tree.* The tree $\mathcal{T}_{\text{prod}}$ is a persistent augmented balanced binary search tree that stores all points in $P$ in order of increasing $y$-coordinate.

Each leaf node $v$ in $T_{\text{prod}}$ stores exactly one point $p = \text{point}[v] \in P$, as well as a boolean field $\text{mark}[v]$. Initially, field $\text{mark}[v]$ is set to false for all leaves in the tree.

Let $v$ be a node in $T_{\text{prod}}$. Let $P(v)$ denote the set of points that are stored in the subtree of $v$, and let $\text{Marked}(v)$ denote the set of points in $P(v)$ for which field mark is set to `true`. Let $lc[v]$ and $rc[v]$ respectively denote the left and right child of $v$, if these children exist. Node $v$ stores two real numbers $\text{iprod}[v]$ and $\text{sprod}[v]$. If $v$ is leaf node, then iprod is set to $1/(1-\pi(\text{point}[v]))$ and $\text{sprod}[v] = D(\text{point}[v])$. If $v$ is an internal node, these values are defined as follows:

$$\text{sprod}[v] = \sum_{q \in \text{Marked}(v)} D(q) \cdot \frac{1}{\overline{\overline{\pi}}\left(\substack{r \in P(v)\setminus \text{Marked}(v) \\ r_y > q_y}\right)}\,,$$

$$\text{iprod}[v] = \frac{1}{\overline{\overline{\pi}}(P(v) \setminus \text{Marked}(v))}\,.$$

We can compute the values $\text{sprod}[v]$ and $\text{iprod}[v]$ in constant time based on the corresponding fields of the node's children. In particular, we have $\text{sprod}[v] = \text{sprod}[lc[v]] \cdot \text{iprod}[rc[v]] + \text{sprod}[rc[v]]$, and $\text{iprod}[v] = \text{iprod}[lc[v]] \cdot \text{iprod}[rc[v]]$.

The product tree supports two operations: $\text{Addmark}(p)$ and $\text{Query}(p)$. For a given point $p$, $\text{Addmark}(p)$ sets the mark field of the corresponding tree leaf to `true`, and updates accordingly values sprod and iprod for every ancestor of this leaf. Let $\text{Marked}(\mathcal{T}_{\text{prod}})$ denote the set of leaves/points in the tree that are currently marked. Given a point $p$, operation $\text{Query}(p)$ returns the following quantity:

$$\text{sprod}[\mathcal{T}_{\text{prod}}](p) = \sum_{q \in P_y^+(p) \cap \text{Marked}(\mathcal{T}_{\text{prod}})} D(q) \cdot \frac{1}{\overline{\overline{\pi}}(P_y^+(p) \setminus \text{Marked}(\mathcal{T}_{\text{prod}}))}\,. \tag{6}$$

Let $p$ be a point in $P$, and suppose that the set of leaves currently marked in $\mathcal{T}_{\text{prod}}$ correspond exactly to those points whose $x$-coordinate is smaller than $p_x$. Then, for this configuration of $\mathcal{T}_{\text{prod}}$, the value returned by $\text{Query}(p)$ is equal to $B(p)$. This observation is the key idea for using the product tree in calculating values $B(p)$. We proceed now with describing how we can efficiently perform operation $\text{Query}(p)$ for any given value $y_o$; first, we split $\mathcal{T}_{\text{prod}}$ into two product trees $\mathcal{T}^-(p)$ and $T^+(p)$ (note that this does not affect $\mathcal{T}_{\text{prod}}$, as it is fully persistent). Tree $\mathcal{T}^+(p)$ stores the points whose $y$-coordinates are strictly larger than $p_y$, and $\mathcal{T}^-(p)$ stores the rest of the points. We then output the value of the field sprod stored at the root of $T^+(p)$; as it can be concluded from the definition of the product tree, this value is equal to the quantity that appears in Eq. (6). Given that the underlying structure of $\mathcal{T}_{\text{prod}}$ is a balanced binary tree, we can perform the described process for Query, but also operation Addmark in $O(\log n)$ time.[1]

[1] We can also extract the value described in Eq. (6) in $O(\log n)$ time without splitting the tree; it suffices to locate the leaf that stores the successor of $p$, and then appropriately process the data stored on the path between this leaf and the root.

To compute values $B(p)$ for every $p \in P$ we work as follows; we first construct $\mathcal{T}_{\text{prod}}$, and we sweep the points in $P$ in order of increasing $x$-coordinate. For each point $p$ that we process, we execute operation Addmark$(p)$, and then Query$(p)$. As outlined in our description for operation Query, at the moment that we process $p$ the value returned from Query$(p)$ is equal to $B(p)$.

Initializing the tree $\mathcal{T}_{\text{prod}}$ and sorting the points on increasing $x$-coordinate take $O(n \log n)$ time in total. The operations that we perform on the tree require $O(\log n)$ time for each point. It follows that we can compute $B(p)$, for all points $p \in P$, in $O(n \log n)$ time in total. The product tree requires $O(n)$ storage, and the lemma follows.

Computing $C(p)$ for every $p \in P$ is quite simpler. Recall that

$$C(p) = p_x p_y \cdot \pi(p) \cdot \overline{\pi}(P_x^+(p) \cup P_y^+(p)) \ .$$

From the latter quantity, we see that computing $C(p)$ fundamentally boils down to calculating $\overline{\pi}(P_x^+(p) \cup P_y^+(p))$. We can do that using a similar approach as with values $B(p)$, except that the data structure $\mathcal{T}$ that we use supports a query operation Query$(p)$ that returns the value $\overline{\pi}(P_y^+(p) \setminus \text{Marked}(\mathcal{T}))$. This structure is simpler than the product tree described above, and it is straightforward to prove that all of its operation can be supported in $O(\log n)$ time. □

From Lemmas 1 and 2, and based on the rest of the analysis that we present in this section, we conclude that value $\mathbb{E}[\max_{p \in S} p_x \cdot \max_{q \in S} q_y]$ can be computed for the Bernoulli distribution in $O(n \log n)$ time. Using this together with Eq. (1), we obtain:

**Theorem 3.** *Let $P$ be a set of $n$ points in $\mathbb{R}^2$, and let $S \subseteq P$ be a random subset selected according to the Bernoulli distribution. We can compute $\mathbb{E}[\text{vol}(\mathcal{BB}(S))]$ in $O(n \log n)$ time.*

In the appendix, we describe an algorithm that computes $\mathbb{E}[\text{vol}(\mathcal{BB}(S))]$ under the Bernoulli distribution when $P$ is a $d$-dimensional point set. This algorithm runs in $O(n^{d-1} \log n)$ time, and uses some of the techniques that we present for the $2D$ version of this problem. For completeness, we also provide in the appendix a description of a $O(n^d \log n)$ algorithm for solving the $d$-dimensional version when subsets of points are selected according to the fixed-size distribution.

## 4 The Volume of the Convex Hull

We next examine the problem of computing the expected volume of the convex hull for a sample of points $S$ selected at random from a point set $P \subset \mathbb{R}^d$. Let $O$ denote the origin in $\mathbb{R}^d$, and without loss of generality we consider that $O$ falls *outside* the convex hull of the entire point set $P$. Let $S$ be a subset of points in $P$, and let $\mathcal{CH}(S)$ denote the convex hull of $S$. Let $F$ be a $(d-1)$-dimensional facet of $\mathcal{CH}(S)$. We use SX$(F)$ to represent the $d$-dimensional simplex whose vertices consist of $O$ and the vertices of $F$. We say that $F$ is a *lower* facet of $\mathcal{CH}(S)$ if SX$(F)$ does not overlap with the $d$-dimensional volume in the interior of $\mathcal{CH}(S)$. Otherwise, we say that $F$ is an *upper* facet of $\mathcal{CH}(S)$. Let UP$(S)$ and LW$(S)$ denote the sets of the upper and lower facets of $\mathcal{CH}(S)$ respectively. We get the next lemma.

**Lemma 4.** *Let* vol(SX$(X)$) *denote the volume of the $d$-dimensional simplex* SX$(X)$*. We have*

$$\text{vol}(\mathcal{CH}(S)) = \sum_{F \in \text{UP}(S)} \text{vol}(\text{SX}(F)) - \sum_{K \in \text{LW}(S)} \text{vol}(\text{SX}(K)). \tag{7}$$

*Proof.* Let $\mathcal{TCH}(S, O)$ denote the union of all the line segments $\overline{Op}$ where $p$ is a point on a $k$-dimensional facet of $\mathcal{CH}(S)$ such that $k < d-1$. Clearly, the $d$-dimensional volume of $\mathcal{TCH}(S, O)$ is zero. To prove the lemma, it suffices to show that a) every point in the interior of $\mathcal{CH}(S) - \mathcal{TCH}(S, O)$ appears in the interior of exactly one simplex SX$(F)$ such that $F \in \text{UP}(S)$, and b) every point in the interior of a simplex SX$(K)$ such that $K \in \text{LW}(S)$ appears in the interior of exactly one simplex SX$(F)$ such that $F \in \text{UP}(S)$. We next prove argument (a), the proof for argument (b) is similar. Let $p$ be a point in the interior of $\mathcal{CH}(S)$ such that $p \notin \mathcal{TCH}(S, O)$. Consider the line $\ell$ that connects $p$ and $O$. Since $\mathcal{CH}(S)$ is a convex set and $p$ lies in the interior of $\mathcal{CH}(S)$ and $p \notin \mathcal{TCH}(S, O)$, then $\ell$ intersects the boundary of $\mathcal{CH}(S)$ in exactly two points; one of these two intersections lies in the interior of a lower facet of the hull, and the other intersection lies in the interior of an upper facet $F$. By construction, $p$ lies inside SX$(F)$ and this simplex is unique for $p$. □

Equation (7) expresses the volume of $\mathcal{CH}(S)$ as a sum of signed volumes, an approach which is similar to the ones of Lasserre [11] and Lawrence [3]. Let $S$ be a randomly selected subset of point set $P$. Given the above description, and due to the linearity property of expectation, we can express the expected volume of the convex hull $\mathcal{CH}(S)$ of $S$ as

$$\mathbb{E}[\text{vol}(\mathcal{CH}(S))] =$$

$$\mathbb{E}[\sum_{F \in \text{UP}(S)} \text{vol}(\text{SX}(F)) - \sum_{K \in \text{LW}(S)} \text{vol}(\text{SX}(K))] =$$

$$\sum_{\substack{Z \subseteq P \\ |Z| = d}} \text{vol}(\text{SX}(F_Z)) \cdot (\mathbb{P}[F_Z \in \text{UP}(S)] - \mathbb{P}[F_Z \in \text{LW}(S)]) \tag{8}$$

where $F_Z$ is the facet defined by the points in $Z$. The volume vol$(\text{SX}(F_Z))$ is equal to $|\det(\text{SX}(F_Z))|/d!$ , where $\det(X)$ is the Cayley-Menger determinant of simplex $X$, that is $\det(X) = \det(v_2 - v_1, .., v_{d+1} - v_1)$, where $v_1, .., v_{d+1}$ are the vertices of $X$. Since we consider that $d$ is constant, we can compute the volume of a simplex in constant time, and thus we can compute the volumes of all simplices that appear in Equation (8) in $O(n^d)$ time in total. Hence, to calculate the expected volume of $\mathcal{CH}(S)$, it remains to compute for every subset $Z$ of $d$ points in $P$ the probabilities $\mathbb{P}[F_Z \in \text{UP}(S)]$ and $\mathbb{P}[F_Z \in \text{LW}(S)]$. We next describe how to compute $\mathbb{P}[F_Z \in \text{UP}(S)]$ for all required subsets $Z$ both in the fixed-size and the Bernoulli model. The probabilities $\mathbb{P}[F_Z \in \text{LW}(S)]$ can be computed in a similar manner.

**Fixed-Size Distribution.** We first consider the case where $S$ is a set of $s$ points selected according to the fixed-size distribution. In this case, we have $\mathbb{P}[F_Z \in \text{UP}(S)] = \binom{n_Z}{s-d}/\binom{n}{s}$, where $n_Z$ is the number of points that appear on the same side of (the hyperplane defining) $F_Z$ as the origin. Next, we show how to compute the values $n_Z$, for all subsets $Z$ of $d$ points in $O(n^d \log n)$ time. It then follows that we can compute $\mathbb{E}[\text{vol}(CH(S))]$ in $O(n^d \log n)$ time as well. We fix a set $Z' \subset P$ of $d-1$ points, and consider all hyperplanes defined by $Z'$, and one additional point from $P$. We order the hyperplanes by radially sorting the points in $P \setminus Z'$ "around" $Z'$. Each hyperplane defines a single candidate facet $F_{Z' \cup \{p\}}$, and it is easy to maintain the number of points that lie on the same side of this hyperplane as the origin. It follows that we can compute all values $n_{Z' \cup \{p\}}$, for all points $p \in P \setminus Z'$ in $O(n \log n)$ time. Since we have to consider $\binom{n}{d-1} = O(n^{d-1})$ sets $Z'$ the total running time for computing $\mathbb{E}[\text{vol}(\mathcal{CH}(S))]$ is $O(n^d \log n)$.

In case $P$ is a set of points in $\mathbb{R}^2$ we can compute the radial order of the points in $P \setminus \{p\}$ around $p$, for every point $p \in P$, in only $O(n^2)$ time in total, while still using $O(n)$ space [16]. It then follows we can also compute $\mathbb{E}[\text{vol}(\mathcal{CH}(S))]$ in $O(n^2)$ time. We obtain the following result.

**Theorem 5.** *Let $P$ be a set of $n$ points in $\mathbb{R}^d$, and let $S \subseteq P$ be a sample selected in the fixed-size model. We can compute $\mathbb{E}[\text{vol}(\mathcal{CH}(S))]$ in $O(n^d \log n)$ time, or $O(n^2)$ time when $d = 2$.*

**Bernoulli Distribution.** For the case where the point set $S$ is selected according to the Bernoulli distribution, we have that $\mathbb{P}[F_Z \in \text{UP}(S)] = \pi(Z) \cdot \overline{\pi}(CO_Z)$, where $CO_Z$ denotes the set of points that lie on a different side of $F_Z$ than the origin, that is the points from $P$ on the "wrong side" of $F_Z$. We can compute $CO_Z$ using a similar approach as with computing values $n_Z$ in the fixed-size distribution. We then obtain the following result.

**Theorem 6.** *Let $P$ be a set of $n$ points in $\mathbb{R}^d$, and let $S \subseteq P$ be a sample selected in the Bernoulli model. We can compute $\mathbb{E}[\mathrm{vol}(\mathcal{CH}(S))]$ in $O(n^d \log n)$ time, or $O(n^2)$ time when $d = 2$.*

## 5 The Mean Distance from the Centroid

Let $S$ be a set of points in $\mathbb{R}^d$, and let $c(S) = \frac{1}{|S|}\sum_{p\in S} p$ denote the *centroid* of $S$, that is, the point representing the coordinate-wise average of the points in $S$. Let $\delta$ be a distance measure in $\mathbb{R}^d$. The *mean centroid distance* based on $\delta$ is equal to the mean value of $\delta$ between any point in $S$ and $c(S)$, that is $\mathrm{CD}(S) = \frac{1}{|S|}\sum_{p\in S}\delta(p, c(S))$. Next, we consider the case where $\delta$ is the squared Euclidean distance. For this case, we obtain the following result.

**Theorem 7.** *Let $P$ be a set of $n$ points in $\mathbb{R}^d$, and let $S \subseteq P$ be a random subset of $s$ points, selected using the fixed-size distribution. We can compute $\mathbb{E}[\mathrm{CD}(S)]$ and $\mathbb{V}[\mathrm{CD}(S)]$ for all subset sizes $s \le n$ in $O(n)$ time in total.*

*Proof.* The expected value of the mean centroid distance for our sample $S$ is

$$\mathbb{E}[\mathrm{CD}(S)] = \mathbb{E}\left[\frac{1}{s}\sum_{p\in S}\sum_{i=1}^{d}(c(S)_i - p_i)^2\right] = \frac{1}{s}\sum_{i=1}^{d}\mathbb{E}\left[\sum_{p\in S}(c(S)_i - p_i)^2\right].$$

The expected value in the last quantity can be expanded as follows:

$$\begin{aligned}
\mathbb{E}\left[\sum_{p\in S}(c(S)_i - p_i)^2\right] &= \mathbb{E}\left[\sum_{p\in S}\left(\left(\sum_{q\in S}\frac{q_i}{s}\right) - p_i\right)^2\right] \\
&= \mathbb{E}\left[\sum_{p\in S}\left(\frac{1}{s}\sum_{q\in S} q_i\right)^2\right] - \frac{2}{s}\mathbb{E}\left[\sum_{r\in S}\sum_{t\in S} r_i\, t_i\right] + \mathbb{E}\left[\sum_{m\in S} m_i^2\right] \\
&= \frac{1}{s}\mathbb{E}\left[\sum_{p\in S}\sum_{q\in S} p_i\, q_i\right] - \frac{2}{s}\mathbb{E}\left[\sum_{r\in S}\sum_{t\in S} r_i\, t_i\right] + \mathbb{E}\left[\sum_{m\in S} m_i^2\right]. \qquad (9)
\end{aligned}$$

Notice that

$$\sum_{r\in S}\sum_{t\in S} r_i\, t_i = 2\sum_{\substack{t,r\in P\\ t\ne r}} r_i\, t_i + \sum_{m\in P} m_i^2\,.$$

Also, the probability that two points $p$ and $q$ are included in $S$ is $\binom{n-2}{s-2}/\binom{n}{s}$, while the probability that a single point $m$ appears in $S$ is $\binom{n-1}{s-1}/\binom{n}{s}$. Based on these observations, we can expand

Equation 9 as follows:

$$\begin{aligned}
&\frac{1}{s}\mathbb{E}\left[\sum_{p\in S}\sum_{q\in S} p_i\, q_i\right] - \frac{2}{s}\mathbb{E}\left[\sum_{r\in S}\sum_{t\in S} r_i\, t_i\right] + \mathbb{E}\left[\sum_{m\in S} m_i^2\right] \\
&= -\frac{2}{s}\sum_{\substack{p,q\in P\\ p\neq q}} p_i\, q_i \frac{\binom{n-2}{s-2}}{\binom{n}{s}} + \left(1-\frac{1}{s}\right)\sum_{m\in P} m_i^2 \frac{\binom{n-1}{s-1}}{\binom{n}{s}} \\
&= -\frac{2\binom{n-2}{s-2}}{s\binom{n}{s}}\sum_{\substack{p,q\in P\\ p\neq q}} p_i\, q_i + \left(1-\frac{1}{s}\right)\frac{\binom{n-1}{s-1}}{\binom{n}{s}}\sum_{m\in P} m_i^2. \qquad (10)
\end{aligned}$$

Observe that $\sum_{\substack{p,q\in P\\ p\neq q}} p_i\, q_i = \sum_{p\in P} p_i(-p_i + \sum_{q\in P} q_i)$, and thus we can compute this term in $O(n)$ time. Clearly, the other term in Eq. (10) can also be evaluated in $O(n)$ time. It follows that we can compute $\mathbb{E}\Big[\sum_{p\in S}(c(S)_i - p_i)^2\Big]$, and thus $\mathbb{E}[\mathrm{CD}(S)]$, in $O(n)$ time as well.

For the variance $\mathbb{V}[\mathrm{CD}(S)]$ we use that $\mathbb{V}[X] = \mathbb{E}[X^2] - (\mathbb{E}[X])^2$, and thus we are interested in computing

$$\begin{aligned}
\mathbb{E}[\mathrm{CD}^2(S)] &= \mathbb{E}\left[\frac{1}{s^2}\left(\sum_{p\in S}\sum_{i=1}^{d}(c(S)_i - p_i)^2\right)^2\right] = \frac{1}{s^2}\sum_{i=1}^{d}\sum_{j=1}^{d}\mathbb{E}\left[\sum_{p\in S}\sum_{q\in S}(c(S)_i - p_i)^2\,(c(S)_j - q_j)^2\right] \\
&= \frac{1}{s^2}\sum_{i=1}^{d}\sum_{j=1}^{d}\left(2\,\mathbb{E}\left[\sum_{\substack{p,q\in S\\ p\neq q}}(c(S)_i - p_i)^2\,(c(S)_j - q_j)^2\right] + \mathbb{E}\left[\sum_{p\in S}(c(S)_i - p_i)^2(c(S)_j - p_j)^2\right]\right). \qquad (11)
\end{aligned}$$

Next, we describe how to compute $\mathbb{E}\Big[\sum_{p\neq q\in S}(c(S)_i - p_i)^2\,(c(S)_j - q_j)^2\Big]$. The second term in Eq. 11, in which $p = q$, can be handled analogously. We use that $(c(S)_i - p_i)^2 = \left(\frac{1}{s}\sum_{r\in S} r_i\right)^2 + p_i^2 - 2p_i\frac{1}{s}\sum_{r\in S} r_i$ and get

$$\begin{aligned}
\mathbb{E}\left[\sum_{\substack{p,q\in S\\ p\neq q}}(c(S)_i - p_i)^2(c(S)_j - q_j)^2\right] = \mathbb{E}\Bigg[\sum_{\substack{p,q\in S\\ p\neq q}}&\left(\frac{1}{s^2}\left(\sum_{r\in S} r_i\right)^2 + p_i^2 - 2p_i\frac{1}{s}\sum_{t\in S} t_i\right) \\
&\left(\frac{1}{s^2}\left(\sum_{m\in S} m_j\right)^2 + q_j^2 - 2q_j\frac{1}{s}\sum_{\ell\in S}\ell_j\right)\Bigg]. \qquad (12)
\end{aligned}$$

We can write this term as the sum of nine terms of the form $\mathbb{E}[\sum_{p\neq q\in P} A(p)B(q)]$, for some functions $A$ and $B$. Next, we focus on the most complicated one, and show that we can evaluate this term in $O(n)$ time. The remaining terms can be handled similarly. We have

$$\mathbb{E}\left[\sum_{\substack{p,q\in S\\ p\neq q}}\left(\sum_{r\in S} r_i\right)^2\left(\sum_{m\in S} m_j\right)^2\right] = \frac{s(s-1)}{2}\cdot\mathbb{E}\left[\left(\sum_{r\in S} r_i\right)^2\left(\sum_{m\in S} m_j\right)^2\right]. \qquad (13)$$

We can expand the expected value that appears in the last quantity as follows :

$$\mathbb{E}\left[\left(\sum_{r\in S} r_i\right)^2 \left(\sum_{m\in S} m_j\right)^2\right] = \sum_{r\in P}\sum_{g\in P}\sum_{m\in P}\sum_{t\in P} g_i\, r_i\, m_j\, t_j \cdot \mathbb{P}[g,r,m,t \in S]\ . \tag{14}$$

We can expand this equation further by considering all possible cases for which $g$,$r$,$m$ and $t$ are either pairwise distinct or represent the same element. There can be fifteen different cases: one in which all four variables represent distinct points in $P$, six cases in which there is exactly one pair of variables representing the same point, etc. We continue with a description on how to efficiently compute one of these cases, the case where $r = m$ and $g = t$. The other cases can be handled in a similar manner. In this case we have $\mathbb{P}[g,r,m,t \in S] = \binom{n-2}{s-2}/\binom{n}{s}$ and

$$\sum_{r\in P}\sum_{\substack{t\in P\\ r\neq t}} r_i\, r_j\, t_i\, t_j = \left(\sum_{p\in P} p_i\, p_j\right)^2 - \sum_{p\in P} p_i^2\, p_j^2.$$

We can compute these terms in $O(n)$ time. It then follows that we can evaluate the term in Eq. 14, and thus also the term in Eq. 12 in $O(n)$ time. This leads to an $O(n)$ time algorithm for computing $\mathbb{E}[\mathrm{CD}^2(S)]$.[2]

In all the quantities that we examine above, parameter $s$ appears in binomial coefficients of the form $\binom{n-k}{s-k}$ where $k$ is a constant. We can precompute all of these coefficients, for all naturals $s \le n$, in $O(n)$ time in total. After this preprocessing stage, we can extract the expected value and variance of $\mathrm{CD}(S)$ in constant time for any sample size $s$. □

## 6 The Mean Pairwise Distance

For a set of points $S \in P$, we call the *mean pairwise distance* of $S$ the average Euclidean distance between any pair of points in $S$. We denote this value by $\mathrm{MPD}(S)$. More formally, we have

$$\mathrm{MPD}(S) = \frac{2}{s(s-1)} \sum_{\substack{p,q\in S\\ p\neq q}} \|pq\|\ ,$$

where $\|pq\|$ denotes the Euclidean distance between $p$ and $q$. We next describe exact and approximate algorithms for computing the expectation and variance of $\mathrm{MPD}(S)$ when $S$ is selected according the fixed-size model from a point set $P \in \mathbb{R}^d$.

**Simple Exact Algorithms for the Fixed-size Distribution.** For the fixed-size distribution, we can express the expected value of the MPD as

$$\mathbb{E}[\mathrm{MPD}(S)] = \frac{2}{s(s-1)} \sum_{\substack{p,q\in P\\ p\neq q}} \|pq\| \cdot \mathbb{P}[\{p,q\} \subseteq S]\ , \tag{15}$$

where $\mathbb{P}[\{p,q\} \subseteq S] = \binom{n-2}{s-2}/\binom{n}{s}$. This leads to a simple algorithm that computes $\mathbb{E}[\mathrm{MPD}(S)]$ in $O(n^2)$ time: compute the Euclidean distance for each pair of points $p, q \in P$ and multiply this distance value by the respective probability of choosing $p$ and $q$ in $S$. Notice that, in the same time asymptotically, it is easy to compute $\mathbb{E}[\mathrm{MPD}(S)]$ for each possible subset size $s$. Computing the variance of MPD is slightly more involved, but can be done in $O(n^2)$ time as well.

[2] The straightforward implementation is quadratic in the dimension $d$. However, it is easy to show that we can improve this to a linear dependence by "pushing in" the sum $\sum_{i,j}^d$ until we get terms of the form $\sum_i^d p_i \sum_{j\neq i}^d p_j = \sum_i^d p_i(-p_i + \sum_j^d p_j)$. These can be evaluated in $O(d)$ time.

**Theorem 8.** *Let $P$ be a set of $n$ points in $\mathbb{R}^d$, and let $S \subseteq P$ be a random subset of $s$ points, selected under the fixed-size distribution. For any given natural $s \leq n$, we can compute $\mathbb{E}[\mathrm{MPD}(S)]$ and $\mathbb{V}[\mathrm{MPD}S]$ in constant time after $O(n^2)$ preprocessing time.*

**An $(1-\varepsilon)$-Approximation Algorithm.** Next, we describe how to compute an $(1-\varepsilon)$-approximation for the expectation and variance of $\mathrm{MPD}(S)$ for the fixed-size distribution. Our approximation algorithm uses a distance $\mathrm{MPD}_\varepsilon(S)$ which is at least $(1-\varepsilon)$ times that of value $\mathrm{MPD}(S)$. More specifically, our algorithm uses a *well-separated pair decomposition* (WSPD) [4] of the points in $P$, and runs in $O(n \log n + n/\varepsilon^d)$ time. This decomposition can be described as follows; Let $z$ be a positive real $\geq 1$. A well-separated pair decomposition $\mathcal{W}_z(P)$ of $P$ with respect to $z$ is a partition of the $\binom{n}{2}$ pairs of points into pairs of point sets, such that for every such pair $(A, B) \in \mathcal{W}_z(P)$, each of the sets $A$ and $B$ fits in a ball of radius $r$, and the distance between these balls is at least $zr$ [4]. We call such a pair a *well-separated* pair. We refer to $z$ as the *separation-factor* of $\mathcal{W}_z(P)$. By choosing $z = 4/\varepsilon$, the distance $\delta_{\mathrm{ball}}(A, B)$ between the balls $\mathcal{B}'(A)$ and $\mathcal{B}'(B)$ containing $A$ and $B$ becomes an $(1-\varepsilon)$-approximation of the distance for any pair of points $a' \in \mathcal{B}'(A)$, $b' \in \mathcal{B}'(B)$. Based on this observation, we consider the following function for a point set $S \subseteq P$:

$$\mathrm{MPD}_\varepsilon(S) = \frac{2}{s(s-1)} \sum_{p,q \in S} \sum_{\substack{(A,B) \in \mathcal{W}_z(P) \\ p \in A, q \in B}} \delta_{\mathrm{ball}}(A, B)$$

Function $\mathrm{MPD}_\varepsilon(S)$ is an $(1-\varepsilon)$-approximation of $\mathrm{MPD}(S)$, therefore its expectation $\mathbb{E}[\mathrm{MPD}_\varepsilon(S)]$ is an $(1-\varepsilon)$-approximation of $\mathbb{E}[\mathrm{MPD}(S)]$ . We can easily derive that $\mathbb{E}[\mathrm{MPD}_\varepsilon(S)]$ is equal to

$$\mathbb{E}[\mathrm{MPD}_\varepsilon(S)] = \frac{2\binom{n-2}{s-2}}{s(s-1)\binom{n}{s}} \sum_{(A,B) \in \mathcal{W}_z(P)} |A|\,|B| \delta_{\mathrm{ball}}(A, B) \ .$$

From this last quantity, it is straightforward to conclude that computing $\mathbb{E}[\mathrm{MPD}_\varepsilon(S)]$ boils down to constructing a well-separated pair decomposition of the desired separation factor. Using the algorithm of Callahan and Kosaraju [4], we can compute a well-separated pair decomposition of $P$ with separation factor $z$ in $O(n \log n + z^d n)$. Since we set $z = 4/\varepsilon$, the running time of this algorithm becomes $O(n \log n + n/\varepsilon^d)$.

We continue by describing an $(1-\varepsilon)$-approximation algorithm for the variance of $\mathrm{MPD}(S)$. We need the following lemma.

**Lemma 9.** *If $X$ is a $(1-\varepsilon)$-approximation of $Y$, then $\mathbb{V}[X]$ is a $(1-2\varepsilon)$-approximation of $\mathbb{V}[Y]$.*

*Proof.* Let $\alpha$ be the exact value such that $X = (1-\alpha)Y$. Since $X$ is a $(1-\varepsilon)$-approximation of $Y$ we have $0 < \alpha \leq \varepsilon$. We then have

$$(1-2\varepsilon)\mathbb{V}[Y] \leq (1-2\alpha)\mathbb{V}[Y] \leq (1-2\alpha+\alpha^2)\mathbb{V}[Y] = (1-\alpha)^2\mathbb{V}[Y] = \mathbb{V}[(1-\alpha)Y] = \mathbb{V}[X], \quad \text{and}$$

$$\mathbb{V}[X] = \mathbb{V}[(1-\alpha)Y] = (1-\alpha)^2\mathbb{V}[Y] < \mathbb{V}[Y].$$ □

Recall that

$$\mathrm{MPD}_\varepsilon(S) = \frac{2}{s(s-1)} \sum_{p,q \in S} \sum_{\substack{(A,B) \in \mathcal{W}_z(P) \\ p \in A, q \in B}} \delta_{\mathrm{ball}}(A, B) \ ,$$

and that $\mathrm{MPD}_\varepsilon(S)$ provides an $(1-\varepsilon)$-approximation of $\mathrm{MPD}(S)$. Based on Lemma 9, we get that the variance of function $\mathrm{MPD}_{\varepsilon/2}(S)$ is an $(1-\varepsilon)$-approximation of $\mathbb{V}[\mathrm{MPD}(S)]$. Given this observation, we can derive an efficient algorithm that calculates an $(1-\varepsilon)$-approximation of $\mathbb{V}[\text{MPD}(S)]$ in $O(n\log n + n/\varepsilon^d)$ time. The (involved) description of this algorithm appears in the proof of the following theorem.

**Theorem 10.** *Let $P$ be a set of points $n$ in $\mathbb{R}^d$, and let $S$ be a subset of $s$ points in $P$ selected according to the fixed-size distribution. For any given natural $s$, we can compute $(1-\varepsilon)$-approximations for the expected value $\mathbb{E}[\text{MPD}(S)]$ and the variance $\mathbb{V}[\text{MPD}(S)]$ of the mean pairwise distance of $S$ in constant time after $O(n\log n + n/\varepsilon^d)$ preprocessing time.*

*Proof.* In Section 6, we already showed how to compute $\mathbb{E}[\text{MPD}_\varepsilon(S)]$, which is an $(1-\varepsilon)$-approximation of $\mathbb{E}[\text{MPD}(S)]$. Based on Lemma 9, we concluded that $\mathbb{E}[\text{MPD}_{\varepsilon/2}(S)]$ is an $(1-\varepsilon)$-approximation of $\mathbb{E}[\text{MPD}(S)]$. We have $\mathbb{V}[\text{MPD}_{\varepsilon/2}(S)] = \mathbb{E}[\text{MPD}^2_{\varepsilon/2}(S)] - (\mathbb{E}[\text{MPD}_{\varepsilon/2}(S)])^2$, and $\mathbb{E}[\text{MPD}_{\varepsilon/2}(S)]$ can be computed in the same way as $\mathbb{E}[\text{MPD}_\varepsilon(S)]$, in $O(n\log n + n/\varepsilon^d)$ time. Hence, we next focus on computing $\mathbb{E}[\text{MPD}^2_{\varepsilon/2}(S)]$.

To do this, we construct a well-separated pair decomposition $\mathcal{W}_z(P)$ with separation factor $z = 4/(\varepsilon/2) = 8/\varepsilon$. Let $p$ and $q$ be two points in $P$ with $p \neq q$, and let $(A,B)$ be the well-separated pair in $\mathcal{W}_z(P)$ such that $p \in A$ and $q \in B$. We use $\delta_{\text{ball}}(p,q)$ to denote the value $\delta_{\text{ball}}(A,B)$, that is the minimum distance between the ball enclosing $A$ and and the ball enclosing $B$. We have

$$\mathbb{E}[\text{MPD}^2_{\varepsilon/2}(S)] = \frac{4}{s^2(s-1)^2} \sum_{\substack{p,q\in P\\ p\neq q}} \sum_{\substack{r,t\in P\\ r\neq t}} \delta_{\text{ball}}(p,q)\cdot\delta_{\text{ball}}(r,t)\cdot \mathbb{P}[\{p,q,r,t\}\subseteq S].$$

We consider three cases for probability $\mathbb{P}[\{p,q,r,t\}\subseteq S]$, depending on whether set $\{p,q,r,t\}$ consists of four, three, or two distinct elements. Based on that, we expand the last quantity as follows:

$$\mathbb{E}[\text{MPD}^2_{\varepsilon/2}(S)] = \frac{4}{s^2(s-1)^2}\left( \frac{\binom{n-4}{s-4}}{\binom{n}{s}} \sum_{\substack{p,q,r,t\in P\\ p\neq q, r\neq t}} \delta_{\text{ball}}(p,q)\cdot\delta_{\text{ball}}(r,t) \right.$$

$$\left. + \frac{\binom{n-3}{s-3}-\binom{n-4}{s-4}}{\binom{n}{s}} \sum_{\substack{p,q,r\in P\\ p\neq q\neq r}} \delta_{\text{ball}}(p,q)\cdot\delta_{\text{ball}}(p,r) + \frac{\binom{n-2}{s-2}-\binom{n-4}{s-4}}{\binom{n}{s}} \sum_{\substack{p,q\in P\\ p\neq q}} \delta_{\text{ball}}(p,q)^2 \right). \tag{16}$$

For the first sum in Eq. (16) we have

$$\sum_{\substack{p,q,r,t\in P\\ p\neq q, r\neq t}} \delta_{\text{ball}}(p,q)\cdot\delta_{\text{ball}}(r,t) = \frac{s^2(s-1)^2\cdot\binom{n-4}{s-4}}{4\binom{n-2}{s-2}}\cdot \mathbb{E}[\text{MPD}_{\varepsilon/2}(S)]^2 .$$

For the third sum in Eq. (16) we have

$$\sum_{\substack{p,q\in P\\ p\neq q}} \delta_{\text{ball}}(p,q)^2 = \sum_{(A,B)\in\mathcal{W}(P)} \delta^2_{\text{ball}}(A,B)\cdot|A|\cdot|B| .$$

It is easy to show that the resulting two quantities can be calculated in $O(n\log n + n/\varepsilon^d)$ time. For the second sum in Eq. (16) we have

$$\sum_{\substack{p,q,r\in P\\ p\neq q\neq r}} \delta_{\text{ball}}(p,q)\cdot\delta_{\text{ball}}(p,r) = \sum_{p\in P} \sum_{\substack{(A,B)\in\mathcal{W}(P)\\ A\ni p}} \sum_{q\in B} \delta_{\text{ball}}(p,q) \sum_{\substack{(G,C)\in\mathcal{W}(P)\\ G\ni p}} \sum_{\substack{r\in C\\ r\neq q}} \delta_{\text{ball}}(p,r) \tag{17}$$

(18)

Since every pair $(p,r)$ occurs in exactly one well-separated pair $(G,C)$, we have

$$\sum_{\substack{(G,C)\in\mathcal{W}(P)\\ G\ni p}} \sum_{\substack{r\in C\\ r\neq q}} \delta_{\text{ball}}(p,r) = -\delta_{\text{ball}}(p,q) + \sum_{\substack{(G,C)\in\mathcal{W}(P)\\ G\ni p}} \sum_{r\in C} \delta_{\text{ball}}(p,r) \ .$$

Thus, we can expand further the quantity in the right side of Eq. 17 as

$$\begin{aligned}
&\sum_{p\in P} \sum_{\substack{(A,B)\in\mathcal{W}(P)\\ A\ni p}} \sum_{q\in B} \delta_{\text{ball}}(p,q) \left( -\delta_{\text{ball}}(p,q) + \sum_{\substack{(G,C)\in\mathcal{W}(P)\\ G\ni p}} \sum_{r\in C} \delta_{\text{ball}}(p,r) \right) \\
&= \sum_{p\in P} \sum_{\substack{(A,B)\in\mathcal{W}(P)\\ A\ni p}} \left( \sum_{q\in B} \delta_{\text{ball}}(p,q) \sum_{\substack{(G,C)\in\mathcal{W}(P)\\ G\ni p}} \sum_{r\in C} \delta_{\text{ball}}(p,r) - \sum_{q\in B} \delta^2_{\text{ball}}(p,q) \right) \\
&= \sum_{p\in P} \left( \sum_{\substack{(A,B)\in\mathcal{W}(P)\\ A\ni p}} |B|\cdot\delta_{\text{ball}}(A,B) \right)^2 - \sum_{p\in P} \sum_{\substack{(A,B)\in\mathcal{W}(P)\\ A\ni p}} |B|\cdot\delta^2_{\text{ball}}(A,B).
\end{aligned} \tag{19}$$

As it comes to the last quantity, the second double sum can be easily calculated in $O(n/\varepsilon^d)$ time by processing each pair of the decomposition in constant time. It remains to describe how we can efficiently compute the first double sum in the last quantity, that is compute values:

$$\text{SUM}(p) = \sum_{\substack{(A,B)\in\mathcal{W}(P)\\ A\ni p}} |B|\cdot\delta_{\text{ball}}(A,B) \ ,$$

for every point $p\in P$. Let $T$ be the split tree of the well-separated pair decomposition, and let node$[p]$ denote the leaf node in $T$ that represents point $p\in P$. Let $v$ be a node in $T$. We use Anc$[v]$ to denote the set of the ancestors of $v$ in $T$. Let $P[v]$ denote the subset of points in $P$ that are represented by the subtree rooted at $v$. We use ms$[v]$ to indicate the sum of all values $k=|B|$ for which $(P[v],B)\in\mathcal{W}(P)$. We represent the sum $\sum_{u\in\text{Anc}[v]}\text{ms}[u]$ by SUM$[v]$. It is obvious that SUM$(p)$ = SUM[node$[p]$]. Hence, to compute SUM$(p)$ for all points in $P$, it suffices to compute SUM$[v]$ for all nodes in $T$. This can be done in $O(n)$ time by a simple top-to-bottom traversal of $T$, assuming that we have already computed values ms$[v]$ for every node $v$ in the tree. The latter values can be computed easily in $O(n\log n + n/\varepsilon^d)$ time when constructing the decomposition.

Notice that parameter $s$ appears in the above quantities either as a constant factor, or within binomial coefficients of the form $\binom{n-k}{s-k}$, where $k\le 4$. We can precompute all these $O(n)$ values for every possible $s$ in $O(n)$ time, separately from the three sums in Eq. (16). Thus, after this preprocessing stage, for any given $s$ we can plug the appropriate coefficients in Eq. (16) and compute the $(1-\varepsilon)$-approximation of the MPD variance in constant time. □

## 7 Diameter of the Smallest Enclosing Disk

Let $\mathcal{D}(S)$ denote the smallest enclosing disk of the set of points $P$, and let $diam(\mathcal{D}(S))$ denote its diameter. We wish to compute the expected diameter of the smallest enclosing disk of $S$. That is, $\mathbb{E}[diam(\mathcal{D}(S))]$. The smallest enclosing disk of $S$ is determined by either two or three points of this set. Let $\partial D$ denote the circle bounding $D$, and let $D_A$ denote the disk that has the points in $A$ on its boundary. Thus, we have

$$\mathbb{E}[diam(\mathcal{D}(S))] = \sum_{p,q\in P} diam(D_{\{p,q\}})\cdot\mathbb{P}[D_{\{p,q\}}=\mathcal{D}(S)] + \sum_{p,q,t\in P} diam(D_{\{p,q,t\}})\cdot\mathbb{P}[D_{\{p,q,t\}}=\mathcal{D}(S)].$$

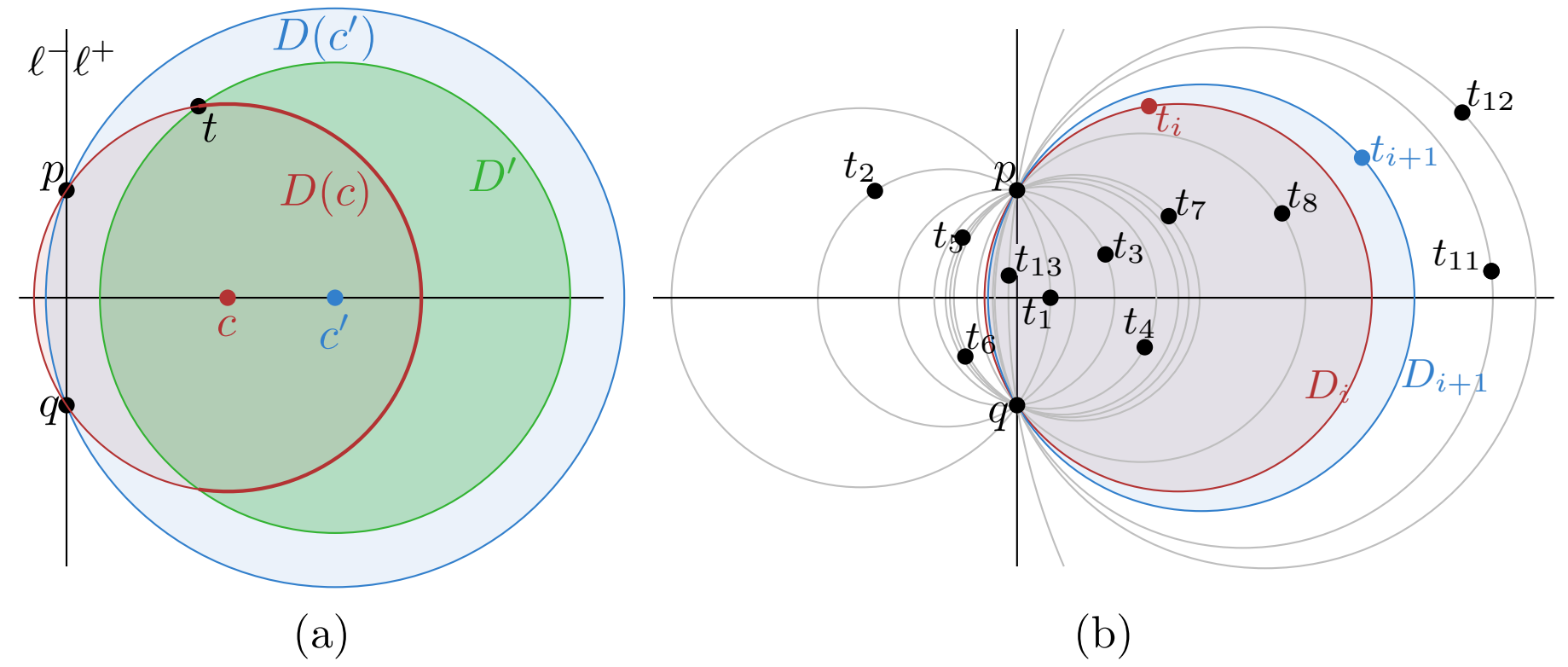

Fig. 1: (a) The disks $D(c)$, $D(c')$, and $D'$ from Lemma 11. It follows that $D(c) \cap \ell^+ \subseteq D(c') \cap \ell^+$. (b) The points $t_1, .., t_{n-2}$. The set of points in $D_{i+1}$ differs by exactly one point from $P \cap D_i$.

We can easily compute the expected diameter for all disks defined by two points in $O(n^3)$ time, so we focus on the case that the smallest enclosing disk is defined by three points, say $p$, $q$, and $t$. We show that we can compute the expected diameter for such disks in $O(n^3 \log n)$ time.

Consider two points $p$ and $q$ and assume without loss of generality that $p$ and $q$ lie on the $y$-axis, and such that the center $m$ of $D_{\{p,q\}}$ corresponds to the origin. Let $\ell$ be the (vertical) line through $p$ and $q$, and let $\ell^-$ and $\ell^+$ denote the left and right half-plane defined by $\ell$, respectively. Observe that all disks $D(c)$ whose boundary contains $p$ and $q$, have their center $c$ on the $x$-axis.

**Lemma 11.** *Let $c$ and $c'$ be two points on the $x$-axis, with $c$ left of $c'$. We have that $D(c) \cap \ell^+$ is contained in $D(c') \cap \ell^+$.*

*Proof.* Let $t \in \ell^+$ be a point on the boundary $\partial D(c)$ of $D(c)$. We show by contradiction that $t \in D(c')$. Since $D(c) \cap \ell^+$ and $D(c') \cap \ell^+$ are convex, the lemma then follows. Let $D'$ be the disk that has center $c'$ and has $t$ on its boundary (See Fig. 1(a)). Since $t \notin D(c')$ the radius $r'$ of $D'$ must be larger than the radius of $D(c')$, and thus $p, q \in D'$. By construction, $t$ is an intersection point of $D'$ and $D(c)$, and $p, q$ lie on the boundary $\partial D(c)$. Since $c'$ lies to the right of $c$, it follows that all points in $D' \cap \partial D(c)$, in particular $p$ and $q$, lie to the right of $t$. Thus, $t$ lies to the left of $p$ and $q$. Contradiction. $\square$

It follows from Lemma 11 that the set of points from $P \cap \ell^+$ that lies in $D(c)$ grows monotonically as we shift $c$ to the right. Analogously, the set of points from $P \cap \ell^-$ shrinks monotonically as we shift $c$ to the right. See Fig. 1(b) for an illustration.

Fix a pair of points $p, q \in P$. Let $T_{pq} = t_1, .., t_{n-2}$ be the points in $P \setminus \{p, q\}$, ordered by increasing $x$-coordinate of the center of $D_{\{p,q,t\}}$, let $D_i = D_{\{p,q,t_i\}}$, and let $S_i = D_i \cap P$. We then have that $S_{i+1} = S_i \cup \{t_{i+1}\}$ or $S_{i+1} = S_i \setminus \{t_{i+1}\}$. We will show that in both distributions we can compute compute $F(p,q) = \sum_{i=1}^{n-2} diam(D_i) \cdot \mathbb{P}[D_i = \mathcal{D}(S)]$ in $O(n)$ time, given the sequence $t_1, .., t_{n-2}$. A straightforward implementation then uses $O(n^3 \log n)$ time: for every pair $p, q \in P$, sort the remaining points in $O(n \log n)$ time and compute $F(p,q)$.

**Algorithms for the Fixed Size and Bernoulli Distributions.** In the fixed-size distribution we have $\mathbb{P}[D_i = \mathcal{D}(S)] = \binom{n-2-|S_i|}{s-2} / \binom{n}{s}$. Since $S_{i+1}$ differs from $S_i$ by at most one point, we can easily maintain this probability in constant time. Computing $diam(D_i)$, for any $i$, also takes constant time, as $D_i$ is defined by only three points. It follows that the total time for computing $F(p,q)$ takes $O(n)$ time in total. In the Bernoulli distribution we have $\mathbb{P}[D_i = \mathcal{D}(S)] = \pi(\{p,q,t_i\}) \cdot \overline{\pi}(P \setminus (\{p,q\} \cup S_i))$. Again, since $S_{i+1}$ differs from $S_i$ by at most one point, we can maintain this probability in constant time, and thus compute $F(p,q)$ in $O(n)$ time. We obtain the following result.

**Theorem 12.** *Given a set of $n$ points in $\mathbb{R}^2$ we can compute the expected diameter of the smallest enclosing disk in $O(n^3 \log n)$ time, in both the Bernoulli and the fixed-size distribution.*

## 8 Experiments and Benchmarks

We implemented three of our algorithms, and evaluated their performance in practice. In particular, we implemented the algorithm that computes the expected $2D$ bounding box volume under the Bernoulli distribution, and the exact and approximate algorithms that compute the expectation and variance of the MPD. As a point of reference, we also implemented the standard heuristic approach [2, 5] currently used in ecological case studies for calculating such values: select a large number of sample point sets according to the examined random distribution (typically a thousand samples), explicitly compute the value of the desired measure on each sample, and extract the mean and variance of these values. We refer to this method as *the sampling method*. We conducted different sets of experiments, where we measured the running time of our implementations for different values of the size $n$ of the input point set $P$, the number of dimensions $d$, and the sample size $s$. For the MPD, we also measured the relative error of the values calculated by the sampling method and our $(1-\varepsilon)$-approximation algorithm.

**Experimental Setup.** We implemented all our algorithms, as well as the naive sampling method, in C++ and using the GNU g++ compiler, version 4.8.3. All experiments were run on a standard desktop computer with a quad core 3.20GHz processor and 8GB of RAM running openSUSE 13.2. To evaluate our algorithms we used both artificial point data, as well as real data representing bird species. The real data set was generated based on data of $n = 9420$ bird species, each of which has ten traits. Since some of these traits admit categorical values (e.g. dietary guild), we had to preprocess the data to map each species to a point in $\mathbb{R}^d$. For this, we used a standard approach in Ecology: we computed a dissimilarity matrix of the species (using the square root of Gower's coefficient [7]), and ran a principle coordinate analysis (PCoA) to produce points of $d = 20$ coordinates. Since the PCoA produces coordinates sorted according to their importance, whenever we performed experiments with this point set in $\mathbb{R}^d$, with $d < 20$, we considered only the first $d$ coordinates of these points. The artificial point set was generated by selecting 9420 points uniformly at random from the 20-dimensional unit cube. In our experiments, for each execution of the sampling method, we fixed the number of selected samples to one thousand.

### 8.1 Experiments on the Mean Pairwise Distance

Next we describe how we evaluated the performance of our algorithms that compute the MPD statistics in the fixed-size model, together with the performance of the standard sampling method. Recall that, unlike the sampling method, our algorithms can compute these statistics for all samples sizes $s \leq n$ for an additional $O(n)$ time. Hence, in all sets of experiments that we describe, the presented running times of our exact and approximation algorithm are for computing the statistics *for all* sample sizes $\leq n$. On the other hand, the presented times for the sampling method indicate the time taken to produce the MPD statistics for a single sample size. In first set of our experiments, we measured the running time of our algorithms as a function of the input size $n$. We considered points in $\mathbb{R}^3$, and we set the sample size to $s = 420$. For the $(1-\varepsilon)$-approximation algorithm, we set $\varepsilon = \frac{1}{2}$. Both for the real and artificial data, we measured the running times of all methods on point sets with size $n = 420 + 500k$, where $k$ ranges from 0 to 18. The results are shown in Fig. 2. The first thing to note is that, since we fixed both the sample size $s$ and the number of repetitions for the sampling method, its running time is independent of $n$. Still, our algorithms are significantly faster than the naive sampling method, even if their running-time measurements include computing the expectation and variance for $n$ different sample sizes, as opposed to the single sample size of the sampling method. On the artificial data, and for the values of $n$ considered, the $(1-\varepsilon)$-approximation algorithm is slightly slower than the exact algorithm. However, as $n$ increases, the difference becomes smaller. We expect that this is mostly due to the larger constant factors in the analysis of the approximation algorithm. For the real data, and data set sizes up to roughly 4000, both algorithms perform similarly. From that point on, the approximation algorithm is faster than the exact algorithm. This suggests that in the real data there are only few well-separated pairs that the algorithm has to consider than in the artificially generated data. This explanation was confirmed by preliminary measurements on the number of generated well-separated pairs. In the above experiments, we also measured the relative error in the values calculated by the $(1-\varepsilon)$-approximation algorithm and the sampling method. It may seem that setting $\varepsilon = \frac{1}{2}$ for our $(1-\varepsilon)$-approximation algorithm is fairly large. However, for all values of $n$ that we considered, the error for the expected mean pairwise distance was always below 1.5 percent; much less than the fifty percent that the algorithm guarantees. Similarly, the maximum observed error for the variance is roughly three percent on the artificial data, and varies between one and three percent for the real data. The error for the expected MPD made by the sampling method is generally very small (close to zero). For the variance, the error of the sampling method is comparable to that of the $(1-\varepsilon)$-approximation, especially for the real data. There does not seem to be a clear relation between the input size $n$, and the error in the approximation or the sampling method. We illustrate these measurements in Fig. 3.

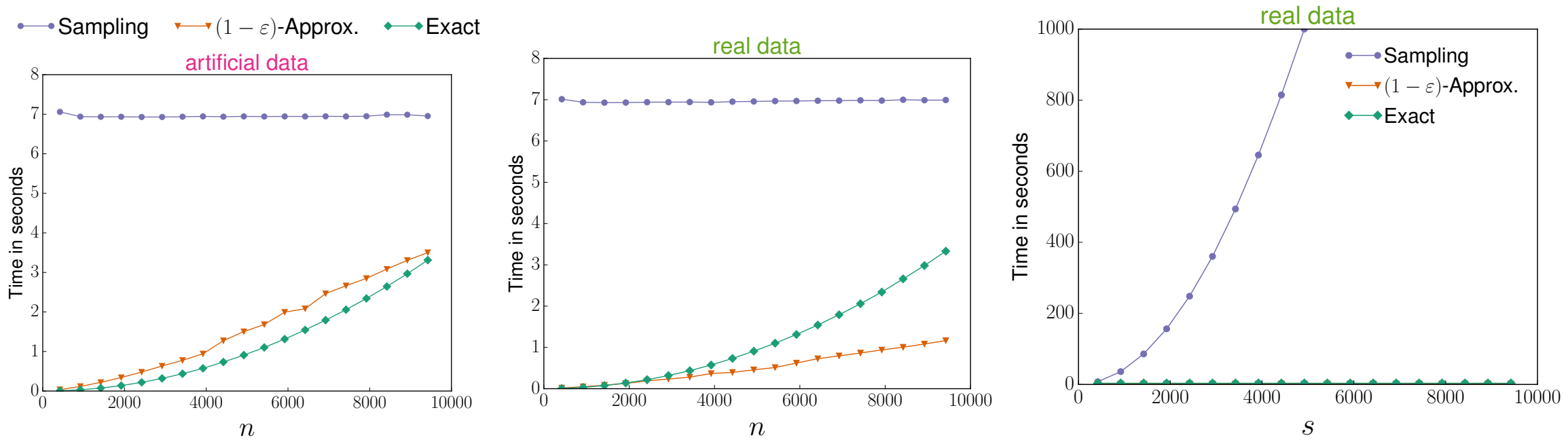

Fig. 2: (left and middle) The running time of our exact MPD algorithm, the $(1-\varepsilon)$-approximation algorithm for $\varepsilon = \frac{1}{2}$, and the sampling method for points in $\mathbb{R}^3$ as a function of $n$. (right) The running time as a function of $s$.

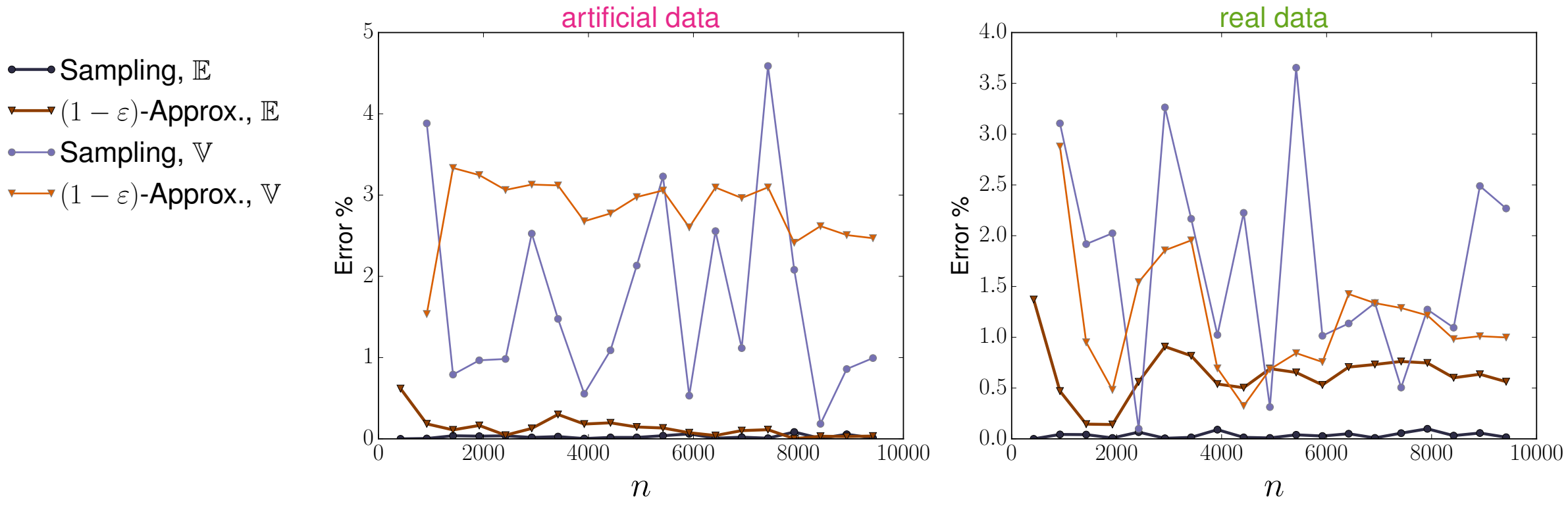

Fig. 3: The observed relative error of our $(1-\varepsilon)$-approximation algorithm and of the sampling method, for different values of $n$. For these experiments, the parameter $\varepsilon$ of the approximation algorithm is set to $\varepsilon = \frac{1}{2}$. However, note that the observed error is much smaller than the guarantee provided by the approximation algorithm.

**Dependency on $s$.** Fixing the sample size $s$ to a small value gives a somewhat unfair advantage to the sampling method. When we vary $s$, we really see the advantages of our algorithms compared to the sampling method. We measured the running times of all methods on the real data, using the full set of 9420 points with $d = 3$ dimensions, and for sample size $s = 420 + 500k$, where $k$ ranges from 0 to 18. The results on the real data are shown in Fig. 2(right). The results on artificial data showed a similar pattern. These results confirm that our algorithms are independent of $s$, whereas the running time of the sampling method slows down significantly as $s$ increases. For $s \approx 4000$ our algorithms are about 200 times faster than the sampling method.

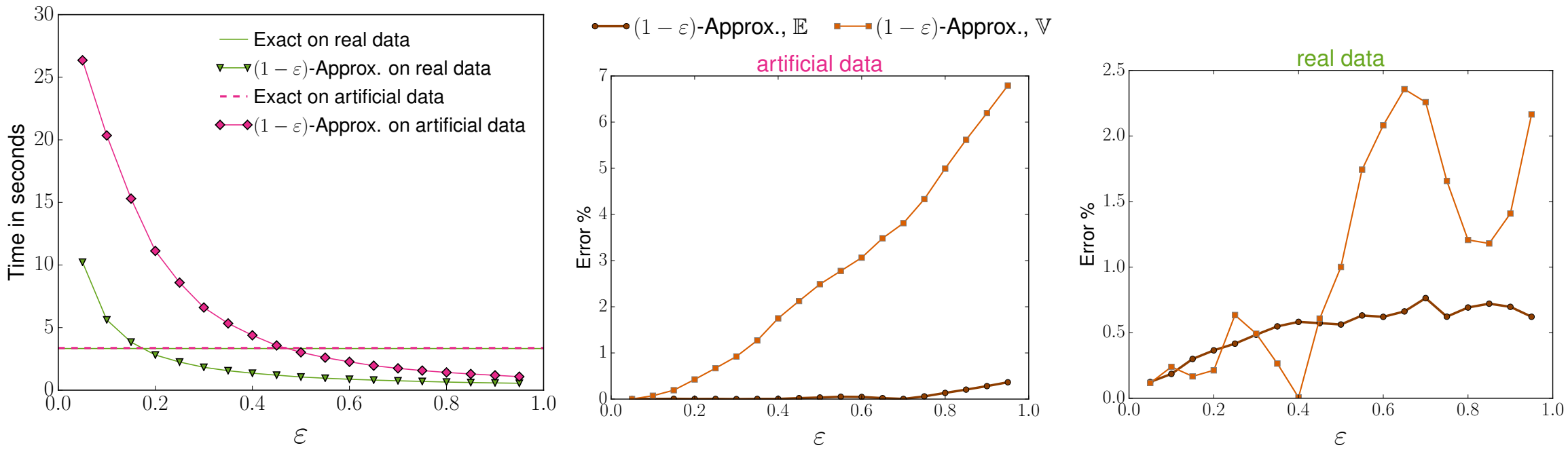

Fig. 4: The running time (left) and observed error (middle and right) of the $(1-\varepsilon)$-approximation algorithm as a function of $\varepsilon$.

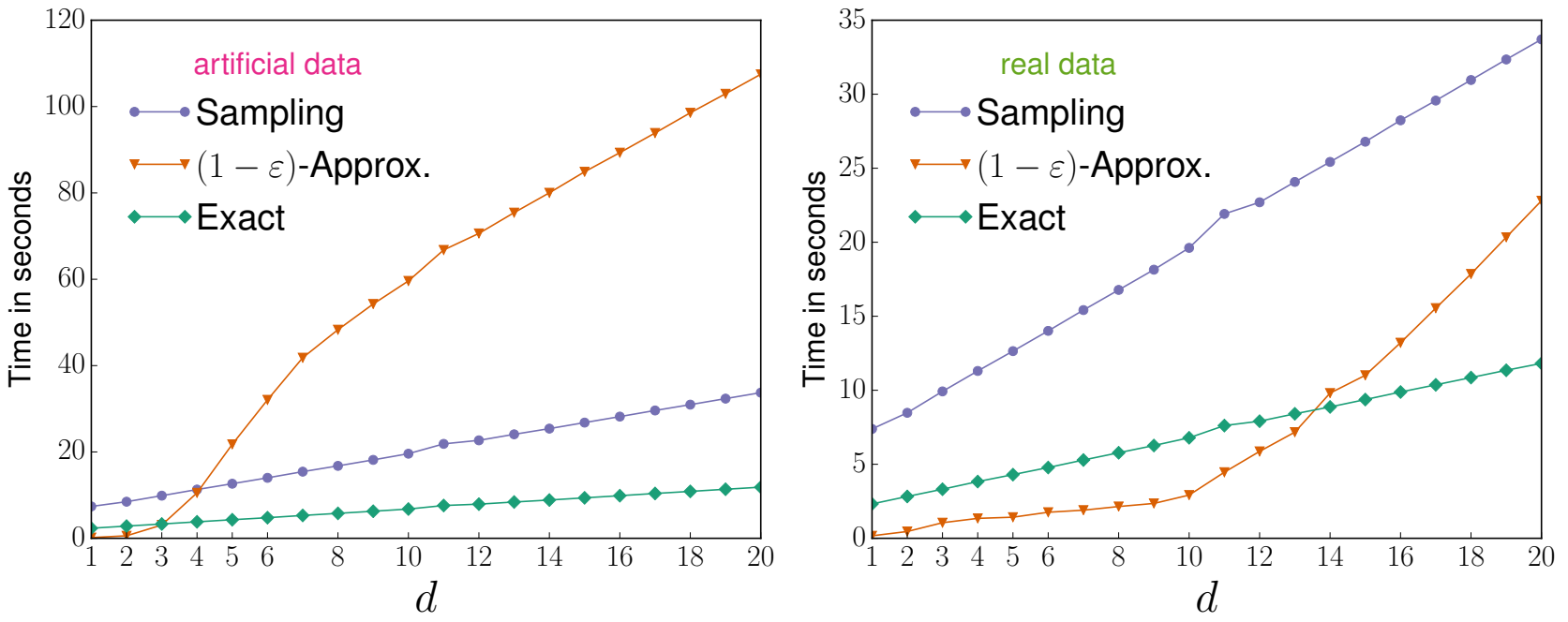

Fig. 5: Running times of the MPD-related methods, as a function of the dimension $d$.

**Dependency on $\varepsilon$.** We also examined how the running time and observed error change for our $(1-\varepsilon)$-approximation algorithm based on $\varepsilon$. We performed experiments both on the real and artificial data, using in each case the full point set. We fixed $d = 3$, and $s = 500$, and we considered $\varepsilon = 0.05 + 0.05k$, with $k$ from 0 to 18. As we see in Fig. 4, the running time for the data set with real species decreases rapidly as we increase $\varepsilon$.

**Dependency on $d$.** Finally, we examined the performance of our algorithms based on $d$. We conducted measurements on both real and artificial data, using the full point sets, and we set $s = 500$ and $\varepsilon = \frac{1}{2}$. The running-time results are shown in Fig. 5. As we could expect from our theoretical analysis, the running times of the sampling and exact algorithms seem to increase smoothly with $d$. On the artificial data, the running time of the $(1-\varepsilon)$-approximation algorithm increases somewhat rapidly for $d \geq 4$. On the real data such an increase seems to show up only for $d \geq 10$. We do see that, on the real data, the approximation algorithm is faster than the exact algorithm up to $d = 13$. To examine this further, investigated the number of well-separated pairs that the approximation algorithm considers. We observed that the increase in the running time was due to the increase on the size of the computed well-separated decomposition. This increase was gradual for the real data, and for the largest examined value of $d$ the size of the decomposition did not exceed 15 percent of the potential maximum of $\binom{n}{2}$ pairs. But, for the artificial data the size of the decomposition rapidly converged to $\binom{n}{2}$, which affected the running time. For every $d$, the observed errors did not exceed three percent (see Fig. 6).

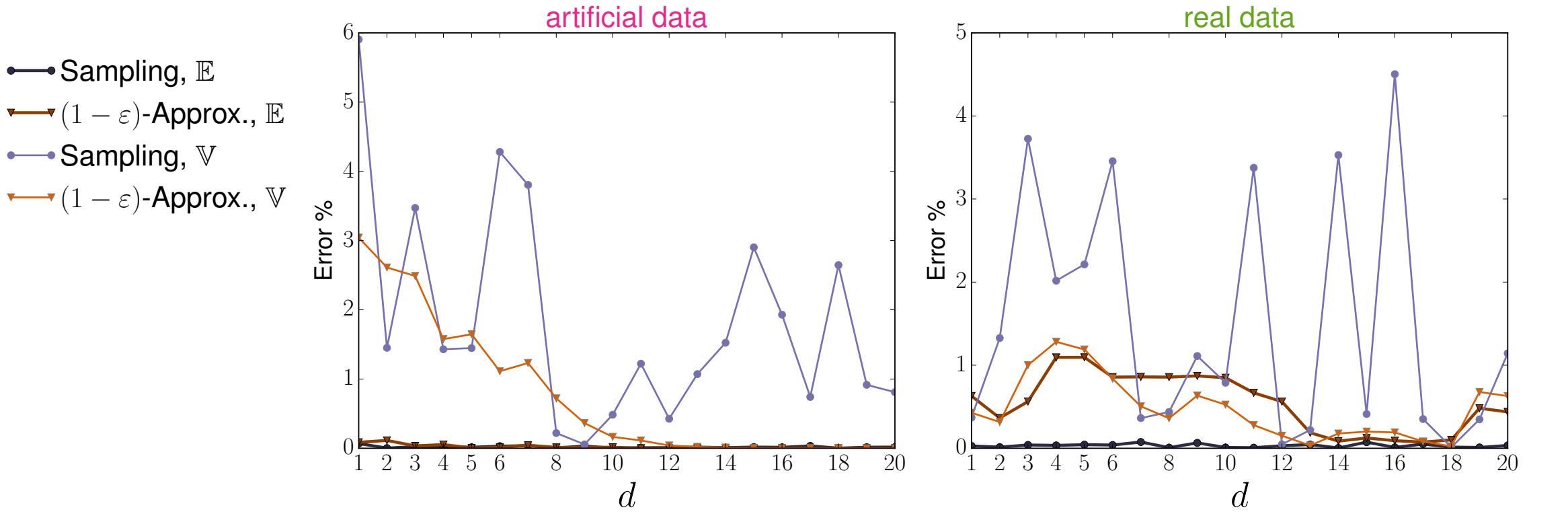

Fig. 6: The observed relative error of our $(1-\varepsilon)$-approximation algorithm and of the sampling method, for different values of $d$.

## 8.2 Experiments on the Bounding Box

We also measured, for different values of $n$, the running time of our algorithm that computes the expected bounding box volume for points in $\mathbb{R}^2$ selected using the Bernoulli distribution. The running time of our method and that of the sampling method on real data is shown in Fig. 7. The results on artificial data are similar. We see that both methods are very fast; they can both process all 9420 points in less than a second. The sampling approach is even slightly faster than our exact algorithm. As with the MPD, the error of the sampling method is close to zero. This could indicate that it may be better to use the standard sampling method on situations where the examined measure can be calculated very fast on a single sample.

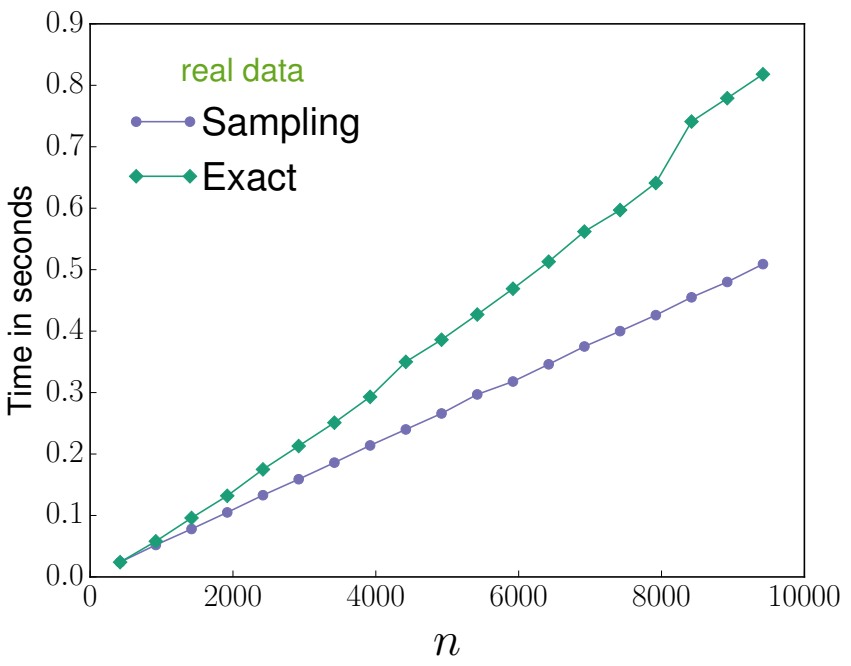

Fig. 7: Running times regarding the 2D bounding box volume in the Bernoulli distribution, for different values of $n$.

# 9 Conclusions and Future Work

We presented polynomial-time algorithms that compute the expectation or variance of popular geometric measures when a subset of points is randomly selected from a universal set $P$ of $n$ points in $\mathbb{R}^d$. We implemented the two algorithms that we designed for computing the statistical moments of the mean pairwise distance for points in $\mathbb{R}^d$, and the bounding box volume for points in $\mathbb{R}^2$. We conducted experiments and we showed that our algorithms are efficient in practice. As part of future research, it would be interesting to study the conditions under which randomized heuristics can produce a good estimation of the mean and variance of an examined measure. For example, suppose that we want to compute a good estimation of the expected convex hull volume of a subset of $s$ points selected from $P$ under the fixed-size model, and that we want to do that by computing the hull volume for a large number of samples of $s$ points. Using Hoeffding's inequality [8] we could decide what is a sufficient number of samples, so that we produce a good estimation with high probability. However, this inequality requires that we provide a good bound for the minimum and maximum value of the measure for a given sample. This leads to the following interesting problem; given a set of points $P$ and a geometric measure $M$, what is the minimum and maximum value that $M$ can take if a subset of points is selected from $P$ under the fixed-size or the Bernoulli model. Löffler and van Kreveld [14] have studied similar problems for imprecise points, that is when the inputs points do not have a fixed position, but they are distributed across given regions. To the best of our knowledge, and especially for the fixed-size model, this problem remains open for most of the measures that we examined in this paper.

## Acknowledgments

We want to thank Vincent Pellissier for introducing the examined problems, and for providing real-world species data for our experiments. We also want to thank Vissarion Fysikopoulos for his contribution in the early stages of this work, regarding the computation of the expected convex hull volume.

This work was supported by the Danish National Research Foundation under grant nr. DNRF84.

## Appendix: The volume of the bounding box in $\mathbb{R}^d$

**Bernoulli Distribution.** Let $p$ be a point in $\mathbb{R}^d$. We use $p_i$ to denote the $i$-th coordinate of this point. Let $S$ be a subset of points in $P$ selected at random according to the Bernoulli distribution. We define

$$\mathrm{XP}(S) = \mathbb{E}\left[\prod_{i=1}^{d} \max_{p \in S} p_i\right].$$

We next describe an algorithm for computing the expected value of the bounding box volume $\mathcal{BB}(S)$ when $S$ is a subset selected according to the Bernoulli distribution. Following a similar analysis as with the 2-dimensional case (see Section 3), it is easy to show that the computation of $\mathbb{E}[\mathcal{BB}(S)]$ for $d$-dimensional point sets boils down to calculating $2^d$ quantities that are fundamentally the same as $\mathrm{XP}(S)$. Therefore, in the rest of this section we describe an algorithm for computing $\mathrm{XP}(S)$ in $O(n^{d-1} \log n)$ time; this can be directly used to derive an algorithm that calculates $\mathbb{E}[\mathcal{BB}(S)]$ in the Bernoulli distribution in $O(2^d n^{d-1} \log n)$ time, which becomes $O(n^{d-1} \log n)$ time in the case that $d$ is a constant.

A brute-force approach for calculating $\mathrm{XP}(S)$ would be to consider all possible subsets of $P$, and for each such subset $S$ compute the maximum coordinates $\max_{p \in S} p_i$ and the probability that $S$ is selected according to the Bernoulli distribution. Of course, this approach would be infeasible since it would require processing all $2^n$ subsets of $P$. Yet, we can design a more efficient approach by considering the following observation; for every $S \subseteq P$, there is a subset $S' \subseteq S$ of at most $d$ points that define the maximum coordinates in this subset. We call such a set $S'$ a *concise set*. Instead of checking explicitly all possible subsets in $P$, we can evaluate $\mathrm{XP}(S)$ by considering only the $O(n^d)$ concise sets in $P$. We next describe how we can do that in $O(n^{d-1}\log n)$ time, without even constructing all these sets explicitly.

Before we continue with our description, we need the following definitions. Let $S$ be a set of $k \le d$ points in $P$. We say that $S$ is a *concise k-set*, or simply a *concise set*, if for every $p \in S$ there exists at least one coordinate $i$ such that $p_i = \max_{q\in S} q_i$. Recall that no pair of points in $P$ share a common coordinate; hence, for each of the $d$ coordinates there is a unique point in a set $S$ that has the maximum value in $S$ for this coordinate. We use $\mathrm{CS}_k(P)$ to denote the set that consists of all concise $k$-sets that are subsets of $P$. We use $\mathrm{CS}(P)$ to denote the union of all sets $\mathrm{CS}_k(P)$, that is the set of all concise sets in $P$. Let $S$ be a concise set. We use $P^+(S)$ to denote the set of all points $p \in P$ such that there exists at least one coordinate $i$ for which $p_i > \max_{q\in S} q_i$. We use $\mathrm{con}(S)$ to denote the following quantity:

$$\mathrm{con}(S) = \prod_{i=1}^{d} \max_{p\in S} p_i \prod_{q\in S} \pi(q) \cdot \overline{\pi}(P^+(S)) \ .$$

We call $\mathrm{con}(S)$ the *contribution* of $S$. Based on these definitions, we can rewrite $\mathrm{XP}(S)$ as follows:

$$\mathrm{XP}(S) = \mathbb{E}\left[\prod_{i=1}^{d} \max_{p\in S} p_i\right] = \sum_{H\in \mathrm{CS}(P)} \prod_{i=1}^{d} \max_{p\in H} \prod_{q\in H} \pi(q) \cdot \overline{\pi}(P^+(H)) = \sum_{H\in \mathrm{CS}(P)} \mathrm{con}(H) \ . \qquad (20)$$

According to the last equation, we can calculate $\mathrm{XP}(S)$ by constructing explicitly all possible concise sets in $P$, and for each such set $R$ compute its contribution. To compute these values efficiently, we break $\mathrm{CS}(P)$ into three parts; the first part contains all concise $k$-sets with $k \le d-2$, the second part contains all concise $(d-1)$-sets, and the third part contains all concise $d$-sets:

$$\sum_{H\in \mathrm{CS}(P)} \mathrm{con}(H) = W_{\le d-2}(P) + W_{d-1}(P) + W_d(P) \,,$$

where

$$W_{\le d-2}(P) = \sum_{k=1}^{d-2} \sum_{H\in \mathrm{CS}_k(P)} \mathrm{con}(H) \,,$$

$$W_{d-1}(P) = \sum_{Q\in \mathrm{CS}_{d-1}(P)} \mathrm{con}(Q) \,, \text{ and}$$

$$W_d(P) = \sum_{R\in \mathrm{CS}_d(P)} \mathrm{con}(R) \,.$$

We next outline how to calculate each of the values $W_{\le d-2}(P)$, $W_{d-1}(P)$, and $W_d(P)$ in $O(n^{d-1}\log n)$ time. It is straightforward to do this for $W_{\le d-2}(P)$; we explicitly construct all concise $k$-sets with $k \le d-2$, and for each such set $R$ determine $\overline{\pi}(P^+(R))$ by naively checking which points in $P$ fall inside $P^+(R)$. There are $O(n^{d-2})$ such concise sets, and it takes $O(n)$ time for each set to determine $P^+(R)$, which leads to $O(n^{d-1})$ time for this process. We continue with the computation of value $W_{d-1}(P)$. We need the next lemma.

**Lemma 13.** *Let $R$ be a concise $k$-set in $P$, and let $Q \subset R$ be a non-empty subset of $R$ that consists of $c$ points. Subset $Q$ is a concise $c$-set, and there exist exactly $\binom{k}{c}$ elements in $\mathrm{CS}_c(P)$ which are subsets of $R$.*

*Proof.* Point set $Q$ is a concise set because each of its points has the maximum value for at least one coordinate $i$ among the points in $R$, and therefore the same point has the maximum value for this coordinate among any subset of $R$. Since every subset of $R$ is a concise set, all subsets of $c$ points in $R$ are elements of $\mathrm{CS}_c(P)$. $\square$

Let $Q$ be a concise $k$-set in $P$, and let $r$ be a natural number greater than $k$. We use $\mathrm{SP}_r(Q)$ to denote all the concise $r$-sets that are supersets of $Q$. Using Lemma 13, we can rewrite $W_{d-1}(P)$ as follows:

$$W_{d-1}(P) = \frac{1}{d-1} \sum_{Q \in \mathrm{CS}_{d-2}(P)} \sum_{R \in \mathrm{SP}_{d-1}(Q)} \mathrm{con}(R)\ . \tag{21}$$

Based on the latter equation, we can calculate $W_{d-1}(P)$ by constructing all concise $(d-2)$-sets, and for each such set $Q$ compute the contributions of the supersets in $\mathrm{SP}_{d-1}(Q)$. We next show how to compute the contributions of all sets in $\mathrm{SP}_{d-1}(Q)$ in $O(n \log n)$ time. Let $S$ be a $k$-concise set with $k < d$. We say that the $i$-th dimension is *non-exclusive* with respect to $S$ if the point $p$ that has the maximum coordinate value in $S$ for this dimension also has the maximum coordinate value in $S$ for another dimension $j \neq i$. We denote the set of non-exclusive dimensions of $S$ by $\mathrm{ND}(S)$. Let $R$ be a concise $(k+1)$-set which is a superset of $S$, and let $q$ be the point such that $R = S \cup \{q\}$. Set $R$ has one less non-exclusive dimension than $S$, which is a dimension where $q$ has a larger coordinate value compared to the points in $S$. Let $i$ be this dimension. We say in this case that point $q$, and respectively superset $R$, *improves* $S$ at dimension $i$. A concise $(d-2)$-set $S$ can have either three or four non-exclusive dimensions, and a concise $(d-1)$-superset of $S$ can improve $S$ in either one or two of these dimensions. Let $\mathrm{SP}^i_{d-1}(S)$ denote the concise $(d-1)$-sets that improve $S$ in dimension $i$ (including also those which improve $S$ in a second dimension), and let $\mathrm{SP}^{\#}_{d-1}(S)$ denote the concise $(d-1)$-sets that improve $S$ in two different dimensions, whichever these dimensions might be. For a concise $(d-2)$-set $Q$, we have

$$\sum_{R \in \mathrm{SP}_{d-1}(Q)} \mathrm{con}(R) = \left( \sum_{i \in \mathrm{ND}(Q)} \sum_{R \in \mathrm{SP}^i_{d-1}(Q)} \mathrm{con}(R) \right) - \sum_{S \in \mathrm{SP}^{\#}_{d-1}(Q)} \mathrm{con}(S)\ . \tag{22}$$

We can compute the contributions of all sets in $\mathrm{SP}^i_{d-1}(Q)$ by extracting all points that improve $Q$ in that dimension, sorting these points in order of increasing $i$-th coordinate, and then computing the contributions of each induced $(d-1)$-superset in constant amortized time based on the contribution of the previous induced set in this order. At the same time, we maintain a sum of the contributions for sets in $\mathrm{SP}^{\#}_{d-1}(Q)$. Each time that we compute the contribution of a concise $(d-1)$-set, we check in $O(d)$ time if this set improves $Q$ in exactly two dimensions and, if this is the case, we add this contribution to the sum that we maintain for $\mathrm{SP}^{\#}_{d-1}(Q)$. The described process requires $O(dn + n \log n)$ time for calculating the contributions of all sets in $\mathrm{SP}_{d-1}(Q)$. Based on Equation 21, we can use this process to compute the sum of these contributions for all $O(n^{d-2})$ concise $(d-2)$-sets, yielding an algorithm that calculates $W_{d-1}(P)$ in $O(n^{d-1} \log n)$ time.

To calculate $\mathrm{XP}(S)$, it remains to compute $W_d(P)$. Using again Lemma 13, we can rewrite $W_d(P)$ as

$$W_d(P) = \frac{2}{d(d-1)} \sum_{Q \in \mathrm{CS}_{d-2}(P)} \sum_{R \in \mathrm{SP}_d(Q)} \mathrm{con}(R)\,. \tag{23}$$

Let $Q$ be a concise $(d-2)$-set in $P$, and let $i$ and $j$ be two non-exclusive dimensions $i, j \in \mathrm{ND}(Q)$. We use $\mathrm{SP}^{ij}_d(Q)$ to denote all the concise $d$-sets that improve $Q$ both in the $i$-th and $j$-th dimension. Based on this definition, we get

$$\sum_{R \in \mathrm{SP}_d(Q)} \mathrm{con}(R) = \sum_{i,j \in \mathrm{ND}(Q)} \sum_{R \in \mathrm{SP}_d^{ij}(Q)} \mathrm{con}(R) \, . \tag{24}$$

With simple arguments we can show that there can be up to four different pairs $i$, $j$ of non-exclusive dimensions for which $\mathrm{SP}_d^{ij}(Q) \neq \emptyset$. Let $i$ and $j$ be such a pair of dimensions, and let $\mathrm{Valid}^{ij}(Q)$ be the set of points in $P \setminus Q$ that improve $Q$ in dimension $i$ and/or dimension $j$ but not in any other of the remaining $d-2$ dimensions. Also, let $P_{ij}^{+}(p,q)$ be the set of points in $r \in P$ such that $r_i > p_i$ and $r_j > q_j$, and let $\max_k(Q)$ be the point with the maximum $k$-th coordinate in $Q$. We get

$$\sum_{R \in \mathrm{SP}_d^{ij}(Q)} \mathrm{con}(R) = \prod_{\substack{k=1 \\ k \neq i,j}}^{d} \max_{q \in Q} q_k \cdot \overline{\pi}(P^{+}(Q) \setminus \mathrm{Valid}^{ij}(Q)) \cdot F(Q,i,j) \ ,$$

where

$$F(Q,i,j) = \sum_{p \in \mathrm{Valid}^{ij}(Q) \cap P_i^{+}(\max_i(Q))} p_i \cdot \pi(p) \cdot \overline{\pi}(P_i^{+}(p) \cap \mathrm{Valid}^{ij}(Q))$$

$$\sum_{r \in \mathrm{Valid}^{ij}(Q) \cap P_i^{+}(\max_j(Q \cup \{p\}))} r_j \cdot \pi(r) \cdot \overline{\pi}(P_j^{+}(q) \cap \mathrm{Valid}^{ij}(Q)) \cdot \frac{1}{\overline{\pi}(P_{ij}^{+}(p,q) \cap \mathrm{Valid}^{ij}(Q))} \, .$$

We can compute the latter quantity in $O(n \log n)$ time; we can extract sets $\mathrm{Valid}^{ij}(Q)$ and $P^{+}(Q)$, and derive values $\prod_{k \neq i,j} \max_{q \in Q} q_k$ and $\overline{\pi}(P^{+}(Q) \setminus \mathrm{Valid}^{ij}(Q))$ in $O(n)$ time. To calculate $F(Q,i,j)$, we need to follow an approach that is very similar to computing value $\sum_{p \in P} B(p)$ for the 2D version of the problem–see Lemma 2 in Section 3. This approach requires using a product tree structure, with the difference that in this case the tree should store one leaf for every point in $\mathrm{Valid}^{ij}(Q) \cap P_j^{+}(\max_j(Q))$, and we consider only queries for each point in $\mathrm{Valid}^{ij}(Q) \cap P_i^{+}(\max_i(Q))$. Applying the described process to every concise $(d-2)$-set yields an algorithm that computes $W_d(P)$ in $O(n^{d-1} \log n)$ time. Together with our analysis for computing $W_{\leq d-2}(p)$ and $W_{d-1}(p)$ this leads to an algorithm that computes $\mathrm{XP}(S)$ in same time asymptotically, and therefore an algorithm that computes the expected bounding box volume in $O(2^d n^{d-1} \log n)$ time, or $O(n^{d-1} \log n)$ time for constant $d$.

**Fixed-size Distribution.** Let $P$ be a set of $n$ points in $\mathbb{R}^d$ and let $S$ be a subset of $P$ which consists of $s$ points and is selected according to the fixed-size distribution. Let $H \subset S$ be the concise $k$-set such that $\max_{p \in H} p_i = \max_{q \in S} q_i$ . For the fixed-size distribution, we redefine the contribution of $H$ as follows:

$$\mathrm{con}(H) = \prod_{i=1}^{d} \max_{p \in H} p_i \cdot \frac{\binom{n-k-|P^{+}(H)|}{s-k}}{\binom{n}{s}} \ , \tag{25}$$

where $|P^{+}(H)|$ is the number of points in $P^{+}(H)$. We see that computing the contribution of $H$ reduces to calculating $|P^{+}(H)|$. The value of $\mathrm{XP}(S)$ is equal to the sum of the contributions of all distinct concise sets in $P$. We can compute $\mathrm{XP}(S)$ under the fixed-size distribution in $O(n^d \log n)$ using a similar approach as with the Bernoulli distribution. For this, we break down $\mathrm{XP}(S)$ into two quantities: the first is $W_{\leq d-1}(P)$, the sum of contributions of all concise $k$-sets with $k \leq d-1$, and the second is $W_d(P)$, the sum of contributions of all concise $d$-sets. Quantity $W_{\leq d-1}(P)$ can be trivially computed by explicitly constructing in $O(n^d \log n)$ time each concise set $H$ with up to $d-1$ points and computing $|P^{+}(H)|$. To evaluate quantity $W_d(P)$, we first construct all concise sets of $d-1$ points. Then, for each such set $H$ we produce all concise $d$-sets that improve $H$, and compute their contibutions in $O(n \log n)$ time. The last step can be performed in a similar way as the computation of $W_{d-1}(P)$ in the Bernoulli model; for each non-exclusive dimension $i$ of $H$, we sort all points in $P^+(H)$ that improve $H$ in increasing order of the $i$-th coordinate, and then compute the contribution of the induced concise sets in constant amortized time.

We can use the above process as a preprocessing stage, and then we can calculate $\mathrm{XP}(S)$ in $O(n)$ time for any given subset size $s$. We can do this as follows; during the process described above, we maintain for every natural $g \leq$ a sum $\mathrm{SUM}_k^g$ where we store values $\max_{p \in H} p_i$ for every concise $k$-set $H$ such that $n-k-|P^+(H)| = g$. Given these $dn$ sums, and given a subset size $s$, we can compute $\mathrm{XP}(S)$ for this subset size in $O(n)$ time using the formula:

$$\mathrm{XP}(S) = \sum_{k=1}^{d} \sum_{g=0}^{n-k} \mathrm{SUM}_k^g \cdot \frac{\binom{g}{s-k}}{\binom{n}{s}} \ . \tag{26}$$

This leads to the following theorem.

**Theorem 14.** *Let $P$ be a set of $n$ points in $\mathbb{R}^d$, and let $S \subseteq P$ be a random sample of $s$ points, selected using the fixed-size distribution. We can compute $\mathbb{E}[\mathcal{BB}(S)]$ for all subset sizes $s \leq n$ in $O(n^d \log n + n^2)$ time in total.*